\documentclass[12pt]{article}

\usepackage{amssymb,amsmath,mathrsfs,epstopdf,slashed,color}
\usepackage{tikz}
\usetikzlibrary{matrix}
\usetikzlibrary{positioning}
\usepackage{ifpdf}
\ifpdf
\usepackage{graphicx}
\usepackage{placeins}
\usepackage{hyperref}
\usepackage{cite}
\else
\usepackage[dvipdfmx]{graphicx}
\usepackage[dvipdfmx]{hyperref}
\usepackage[numbers]{natbib}
\usepackage{hypernat}
\fi

\allowdisplaybreaks[1]

\makeatletter

\@addtoreset{equation}{section}
\makeatother

\DeclareMathOperator*{\Tr}{Tr}
\DeclareMathOperator*{\re}{Re \,}

\begin{document}

\begin{titlepage}

\setcounter{page}{0}
  
\begin{flushright}
 \small
 \normalsize
\end{flushright}

\vskip 3cm
\begin{center}

{\Large \bf Bloch Wave Function for the Periodic Sphaleron Potential and Unsuppressed Baryon and Lepton Number Violating Processes}  

\vskip 2cm
  
{\large S.-H. Henry Tye${}^{1,2}$ and Sam S.C. Wong${}^{1}$}
 
 \vskip 0.6cm

 ${}^1$ Department of Physics and Jockey Club Institute for Advanced Study,
 
  Hong Kong University of Science and Technology, Hong Kong\\
 ${}^2$ Laboratory for Elementary-Particle Physics, Cornell University, Ithaca, NY 14853, USA

 \vskip 0.4cm

Email: \href{mailto:iastye@ust.hk, scswong@ust.hk}{iastye at ust.hk, scswong at ust.hk}

\vskip 1.0cm
  
\abstract{\normalsize For the periodic sphaleron potential in the electroweak theory, we find the one-dimensional time-independent Schr\"{o}dinger equation with the Chern-Simons number as the coordinate, construct the Bloch wave function and determine the corresponding conducting (pass) band structure. We show that the baryon-lepton number violating processes can take place without the exponential tunneling suppression (at zero temperature) at energies around and above the barrier height (sphaleron energy) at 9.0 TeV. Phenomenologically, probable detection of such processes at LHC  is discussed.
 }
  
\vspace{1cm}
\begin{flushleft}
 \today
\end{flushleft}
 
\end{center}
\end{titlepage}

\setcounter{page}{1}
\setcounter{footnote}{0}

\tableofcontents

\parskip=5pt

\section{Introduction}

It is by now well-known that both the baryon number (B) and the lepton numbers (L) are not conserved in the standard electroweak theory because of the presence of anomalies \cite{Adler1969,Bell1969,Hooft1976,Hooft1976a} and instantons \cite{Belavin1975}. The non-Abelian nature of a Yang-Mills theory leads to a topologically non-trivial vacuum structure. An instanton solution (with index $N=1$) in the 4-dimensional Euclidean Yang-Mills field equations yields a tunneling amplitude from one vacuum $\left|n\right\rangle$ to another vacuum $\left|n+1\right\rangle$ of order $\exp (- 2\pi/\alpha)$, where $\alpha = g^2/(4\pi)$ is the gauge coupling; for small couplings, such tunneling is typically exponentially suppressed. Since there are multiple vacua labeled by the number $n$ ($n=-\infty, ... -1, 0, +1,... +\infty$), separated by the same barrier, such a periodic effective potential implies that, in the $SU(3)$ QCD case, the actual QCD vacuum should be described by a Bloch wave function labeled by $\Theta$, i.e.,
$\left|\Theta\right\rangle = \sum_n \exp(i n \Theta) \left|n\right\rangle$ \cite{Jackiw1976,Callan1976a}. Because of the scaling property of the instanton solutions and the running of  the coupling $\alpha_{QCD}$ in QCD, it is difficult to determine the barrier potential in QCD. On the other hand, the electroweak theory has a natural scale, namely the Higgs vacuum expectation value $v$, or equivalently, the W-boson mass $m_W=gv/2$, where $g$ is the $SU(2)$ gauge coupling. The existence of the closely related sphaleron in the electroweak theory \cite{Dashen1974b} was first studied by Manton\cite{Manton1983} and Klinkhamer and Manton\cite{Klinkhamer1984}. Although the sphaleron is not a topological soliton, it does have a Chern-Simons (CS) number (half-integer) and is important to the dynamics in the electroweak theory. The sphaleron energy $E_{sph}$ measures the height of the potential barrier to the baryon- and lepton-number ($B+L$)-violating processes (which conserve the ($B-L$) number).

The sphaleron has been extensively studied because it is likely to play a crucial role in the matter-anti-matter asymmetry in our universe due to the electroweak phase transition in the early universe \cite{Kuzmin1985}, a subject that has been extensively studied (see e.g., Ref \cite{Cohen1993,Rubakov1996,Trodden1999,Morrissey2012,Shaposhnikov1987} for reviews). So it is obviously important to know whether such processes can be observed and studied in the laboratory today. The possibility of sphaleron mediated baryon and lepton number violating processes at high energy colliders has been studied to some extent\cite{Gibbs1995a,Bezrukov2003,Bezrukov2003a,Ringwald2003,Ringwald2003b}. When the energy reach is much lower than the sphaleron energy $E_{sph}=9$ TeV, these baryon number violating processes are exponentially suppressed, by a factor like $\exp (- 4\pi/\alpha_W) \sim 10^{-162}$ where $\alpha_W \simeq 1/30$.  When the parton-parton energy approaches $E_{sph}$, such processes are not as suppressed. However, the result of the analyses done for collider physics so far is somewhat ambiguous. 

Here we like to estimate the ($B+L$)-violating rates in colliders using the periodicity property of the sphaleron potential as well as the presence of the kinetic term for the CS number $n$, which is necessary for a full first quantized treatment of the problem. To our knowledge, the discrete symmetry has not been emphasized in any of the sphaleron studies so far. 
As is well known, the quantum properties of a periodic potential is quite different from that of a single potential barrier. In this paper, we find the corresponding Bloch wave function and the conducting (pass) band (one-dimensional Brillouin zone) structure in the electroweak theory. 

\begin{figure}[h]
 \begin{center}
  \includegraphics[scale=0.7]{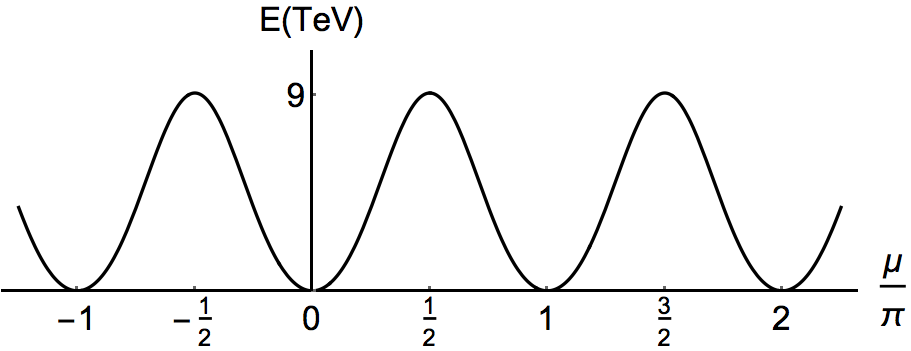}
  \caption{The periodic sphaleron potential $V(Q)$ as a function of the coordinate $Q$ in the electroweak theory. The barrier height is 9 TeV. The dimensionless $\mu=m_WQ$ is related to the Chern-Simons number $n$ via $n=\mu/\pi -\sin(2\mu)/(2\pi)$. The extrema of $V(Q)$ are at $\sin(2\mu)=0$: the minima (vacua) are at integers $n=\mu/\pi= ... , -2, -1, 0, +1, +2, ... $ and the peaks (i.e., the sphaleron) are at $n+1/2$  
\cite{Manton1983}.}\label{fig:intro}
 \end{center}
\end{figure}

Although $n$ always takes an integer value at a vacuum state and a half-integer value at a peak of the potential (i.e., sphaleron solution), it takes continuous values as we move from one vacuum state over the sphaleron potential barrier to the next vacuum state. Since the sphaleron potential has been calculated already \cite{Manton1983,Akiba1988}, we need only to evaluate the ``mass" ${\hat m}$ in the kinetic term ${\hat m}(n) {\dot n}^2/2$ (dot denotes time derivative) to obtain the corresponding one-dimensional time-independent Schr\"{o}dinger equation in which the CS number $n$ is the coordinate to be quantized. For values away from half-integers, $n \ne 0, \pm1/2, \pm 1, ...$, the definition of $n$ is not unique, as the CS current is not gauge-invariant. We find that $\mu/\pi$, instead of the standard choice of $n$, is the most appropriate CS number that takes continuous values (where ${n\pi}= \mu - \sin(2\mu)/2$). Introducing $Q=\mu/m_W$ (so $Q$ has the dimension of a coordinate) we obtain a constant mass $m$ and
\begin{equation}
\label{Sch1}
  \left( - {1 \over 2m}{\partial ^2 \over \partial Q^2} +V(Q) \right) \Psi(Q) = E\Psi(Q).
\end{equation} 
Using the known Higgs vacuum expectation value $v=246$ GeV, $W$ Boson mass $m_W=80$ GeV and the Higgs Boson mass $m_H=125$ GeV, we obtain
\begin{align}
\label{mainresult}
  V(Q) &\simeq 4.75 \mbox{ TeV} \left(1.31\sin^2 (m_WQ)+0.60 \sin^4 (m_WQ) \right), \nonumber\\
  E_{sph} &= \mbox{max}  [V(Q)] =V \left(\frac{\pi}{2 m_W} \right)= 9.11 \mbox{ TeV}, \nonumber \\
   m & =17.1 \mbox{ TeV},  
\end{align}
where the potential $V(Q)$ was obtained by Manton (see Fig. \ref{fig:intro}). 
Determining the value of this mass $m$ is a main result of this paper.  Note that a rescaling of $Q$ rescales $m$, though the physics is unchanged.

To find the mass $m$, we have to make a couple of appropriate changes (gauge rotations) in the existing works. This is necessary because, although the static potential is gauge-invariant, different choices of static gauges tend to yield different masses in the time dependent kinetic term. 
The static sphaleron potential barrier has been calculated in two ways, namely the Manton method \cite{Manton1983} and the method due to Akiba, Kikuchi and Yanagida (AKY) \cite{Akiba1988}. After the necessary modifications just mentioned, we find that the mass $m$ in the Manton approach is a constant, as given in Eq.(\ref{mainresult}), while the AKY mass $m(n)$ diverges (close to a linear divergence) as $n \rightarrow 0$. 
This divergent behavior is close to the simple example of 
${\dot y}^2/(4|y|) + |y|  \rightarrow {\dot x}^2 + x^2$ if $y=x^2$. 
After the redefinition, the constant AKY mass $m=22.5$ TeV (with co-ordinate $Q$) is somewhat larger than the mass $m=17.1$ TeV in the Manton case, while the potentials are close but not the same. So the overall features stay the same.
Clearly a fully time-dependent evaluation of $m$ may improve its value. Fortunately, the present approximation is good enough for our purpose. 

 It was pointed out in \cite{Manton1983,Klinkhamer1992,Kunz1992}  that turning on the $U(1)$ coupling (i.e., Weinberg angle $\sin^2 \theta_W = 0.23$) will lower the sphaleron energy by about a percent. So it is reasonable to use $E_{sph}=9.0$ TeV in phenomenological studies.

Once we have the one-dimensional time-independent Schr\"{o}dinger equation (\ref{Sch1}) with the mass $m$ and the periodic potential $V(Q)$ (\ref{mainresult}), it is straightforward to solve for the Bloch wave function, the conducting (pass) bands, their widths and the gaps between the bands. In Table \ref{tab:bandwidth}, we give the lowest few bands and the ones that are close to the barrier height $E_{sph}=9.11$ TeV. (Because of the higher mass $m$ and higher potential away from the extrema, there are more bands in the AKY estimate.) We see that the first band occurs at about 35 GeV. With energies inside a pass band, the wave function spreads across the whole potential and transmission from one vacuum to another (at different integer $n$) is no longer tunneling suppressed.  However, the width of the bands at low energies are exponentially narrow. Averaging over a few bands and their gaps at low energies, we find that the probability to lie inside a band is exponentially suppressed. This is simply another way to see the tunneling suppression effect. As the energy approaches $E_{sph}$ from below, the widths of the bands become bigger, while the gaps become smaller, so the effect of the pass bands becomes important. This is when the ($B+L$)-violating processes are no longer tunneling suppressed, even when the energy is still a little below the barrier height. 

It is most important to search for these ($B+L$)-violating processes in the laboratory.  Particularly interesting parton (left-handed quarks) scatterings in proton-proton collisions are the $\Delta n =-1$ quark-quark annihilations at close to or above 9 TeV, 
\begin{equation}\label{qqBL-1}
q_L + q_L \rightarrow {\bar l}_e  {\bar l}_{\mu}  {\bar l}_{\tau} {\bar q}{\bar q}{\bar q}{\bar q}{\bar q}{\bar q}{\bar q} + X,
\end{equation}
which results in 3 anti-leptons (one from each family) and 7 anti-quarks plus other particles. The preferred quark content should contain 3 anti-quarks from the second family ($\bar c$ or $\bar s$) and 3 anti-quarks from the third family ($\bar t$ or $\bar b$). The above scattering is an inclusive process, so $X$ (with net $B=L=0$) may include any number of $W^{\pm}$, Z and Higgs bosons, mesons and photons as well as  fermion-anti-fermion pairs. $X$ has a net electric charge to maintain charge conservation of the process.
These events probably look like fireballs. Similarly, one can consider a particular $\Delta n =+1$ quark-quark scattering,
\begin{equation}\label{qqBL+1}
q_L + q_L \rightarrow e^-\mu^- \tau^- bbbcccddduu+ X
\end{equation}
So a single ($B+L$)-violating event can produce 3 positive sign leptons plus 3 ${\bar b}$-quarks ($\Delta n =-1$), or 3 negative sign leptons plus 3 $b$-quarks ($\Delta n =+1$). We expect roughly equal numbers of events for each. A very crude order of magnitude estimate gives $10^{4 \pm 2}$ such events in the coming Large Hadron Collider (LHC) run at 14 TeV proton-proton collisions. Because the fraction of quark-quark scatterings with energies close to or above the sphaleron energy $E_{sph}=9$ TeV is substantially bigger at the 14 TeV than at the 13 TeV, detection may be easier by comparing the two data samples.  If the LHC energy can be raised by a few TeV, the ($B+L$)-violating event rate should increase by more than an order of magnitude.

The rest of the paper is organized as follows. In Sec. 2, we review the properties of the sphaleron and the barrier potential first constructed by Manton \cite{Manton1983}, where the necessary modifications for our purpose are already incorporated. In Sec. 3, we construct the Lagrangian and the Hamiltonian for the Chern-Simons number after finding its kinetic term. After quantization, we obtain the one dimensional Schr\"{o}dinger equation with $Q=\mu/m_W$ (instead of the Chern-Simons number) as the coordinate. It is then straightforward to find the pass band structure in the electroweak theory, which we present in Sec. 6. In Sec. 4, we point out the differences  we have to make in the Manton construction to obtain what we believe to be the correct mass.   In Sec. 5, we review the AKY construction of the potential and find the corresponding mass ${\hat m}(n)$, which diverges as $n\rightarrow 0$. After a transformation to the canonical form, we find that the AKY potential is not as different from the Manton potential as one might initially be led to believe. Sec. 4 and Sec. 5 may be skipped without loss of continuity.  In Sec. 7, we discuss some phenomenology of the ($B+L$)-violating processes. Our main point is the possibility that they may be detected at LHC in the near future. Clearly more theoretical as well as phenomenological studies are warranted. Sec. 8 contains the summary and some remarks. Appendix A contains some details.


\section{Review}

Let us review the basic facts about anomalies, instantons, tunneling and the sphaleron potential. We make some changes to the original work in the study of the sphaleron as this will turn out to be important later.

\subsection{Background}

We start with the $SU(2)$ weak interaction gauge fields $A^a_{\mu}(x)$ coupled to a doublet Higgs field $\Phi(x)$ and left-handed fermion doublets $\Psi^{(i)}_L = (q^{f,a}_L, l^f_L)$, where $f=1,2,3$ is the family index and $a=1,2,3$ is the color index for the quarks. To simplify the discussions, we ignore the $U(1)$ gauge field (i.e., set Weinberg angle $\theta_W=0$) and the right-handed fermions; for our study, a non-zero $\theta_W$ as well as the Higgs coupling to right-handed fermions introducing fermion masses and Yukawa couplings will have only minor changes to the analysis. We shall comment on these minor complications later.

Consider only the simplified version of the standard electroweak model (with $ds^2=dt^2-d\vec{x}^2$ and $\hslash=c=1$),
\begin{align}
  \label{lagran} {\cal L} = -{1\over2}\Tr[F_{\mu\nu}F^{\mu\nu}]+{1\over 2} \left(D_{\mu}\Phi\right)^{\dagger}D^{\mu}\Phi -{\lambda \over4}\left( \Phi^{\dagger}\Phi -v^2 \right) ^2 + i {\bar \Psi}^{(i)}_L \gamma^{\mu}D_{\mu}\Psi^{(i)}_L,
\end{align}
where 
\begin{align}
   F_{\mu\nu}&=F^a_{\mu\nu}{\sigma^a \over 2} = \partial_{\mu}A_{\nu} - \partial_{\nu}A_{\mu}-ig\left[A_{\mu},A_{\nu}\right] ,\nonumber\\
   D_{\mu}\Phi &= \partial_{\mu}\Phi - i g A_{\mu}\Phi, \nonumber\\
   D_{\mu}\Psi^{(i)}_L &= \partial_{\mu}\Psi^{(i)}_L - i g A_{\mu}\Psi^{(i)}_L,
\end{align}
where $A_{\mu}(x)=A_{\mu}^a \sigma^a/2$ and $i=1,2,...,12$ for the 12 doublets of left-handed fermions.

At the classical level, there exist $n_L=12$ ($i=1,2,...,n_L$) globally conserved $U(1)$ currents 
$$J^{(i) \mu}={\bar \Psi}^{(i)}_L \gamma^{\mu}\Psi^{(i)}_L, $$
corresponding to the conservation of the fermion numbers. However, this conservation is broken by the presence of anomaly \cite{Adler1969,Bell1969},
$$\partial_{\mu} J^{(i) \mu} = \frac{g^2}{16 \pi^2} \Tr\left[F_{\mu \nu} \tilde{F}^{\mu \nu}\right]$$
where $\tilde{F}^{\mu \nu}$ is the dual of  ${F}^{\mu \nu}$.
In the presence of instanton solutions in Euclidean space-time\cite{Belavin1975},  
\begin{equation}
 \label{chern} N =\frac{g^2}{16 \pi^2} \int d^4x \Tr\left[F_{\mu \nu} \tilde{F}^{\mu \nu} \right], 
\end{equation} 
where the topological index $N$ takes only integer values. An instanton with value $N$ leads to the tunneling process $\left|n \right\rangle \rightarrow \left|n+N \right\rangle$.
As the change of fermion number $\Delta N^{(i)}$ is the same for each doublet, we have the change in the lepton number $L$ given by
\begin{equation}
\Delta N_{e} = \Delta N_{\mu} = \Delta N_{\tau}= N,
\end{equation}
and the change in the baryon number $B$ given by
\begin{equation}
\Delta B =\frac{1}{3} (3)(3) N=3N,
\end{equation}
since each quark has $B=1/3$ and there are 3 families and 3 colors \cite{Hooft1976}. As a result, the ($B - L$) number is conserved, as $\Delta B -\Delta L=3N-3N=0$. Since the electric charge $Q_e$ is always conserved, we have
\begin{equation}
\Delta (B+L) = 6N, \quad \quad \Delta (B-L) = \Delta Q_e = 0.
\end{equation}
For example, an $e^+e^-$ collision can produce 3 baryons : $e^+e^- \rightarrow e^-\mu^-\tau^-B_c^+ B_c^+B_c^+$ where each baryon $B_c^+$  is a bottom-charm baryon which decays to a nucleon plus mesons, e.g., $B_c^+(bcu) \rightarrow p  \pi^+ \pi^- \pi^0$. 
By comparison, a process in a proton-proton collision $pp \rightarrow e^+ \mu^+ {\bar \nu}_{\tau} \bar n $ can be mediated by an instanton also, though this process is further suppressed by the CKM mixing. At low energies, since an instanton action $S=2 \pi/\alpha_W$ where the weak coupling $\alpha_W \sim 1/29.7$, the tunneling rate goes like $\exp(-2S) \simeq 10^{-162}$, which is totally unobservable.

In this paper, we propose that, under appropriate conditions,  ($B+L$)-violating processes may not be exponentially suppressed. 
It is known that such ($B+L$)-violating processes become much less suppressed at high temperatures \cite{Arnold1987,Arnold1988,Ringwald1988} and it may be responsible for the matter-anti-matter asymmetry of our universe \cite{Kuzmin1985}. Here we restrict ourselves to the zero temperature case and such processes in the laboratory only.

\begin{figure}[h] 
 \begin{center}
  \includegraphics[scale=.5]{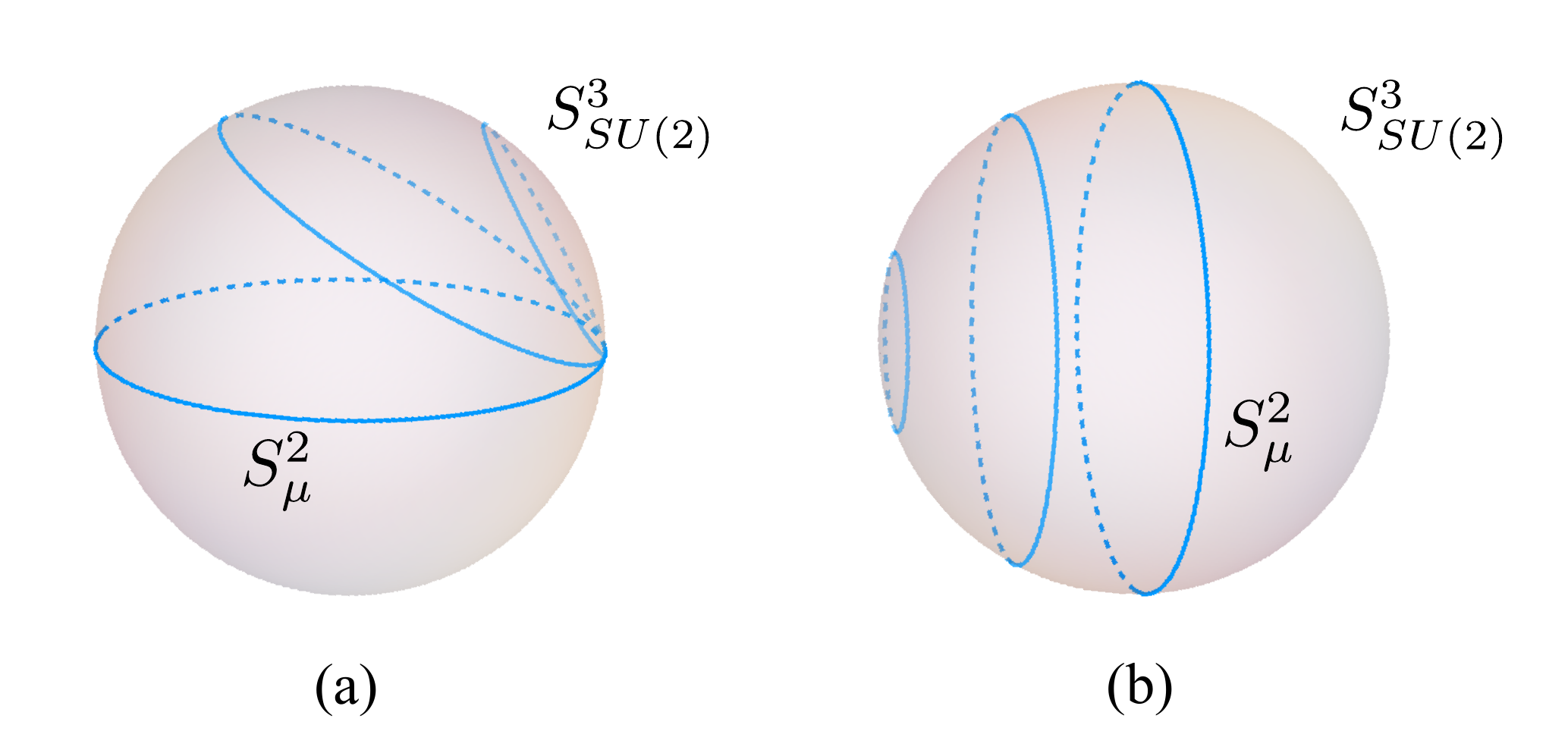}
  \caption{Here the 3-sphere is spanned by the point $p(\mu, \theta, \varphi)$ where $0 \le \mu < \pi$ and the usual polar angles $(\theta, \varphi)$ spans a 2-sphere. The sphaleron corresponds to $\mu=\pi/2$, when the 2-sphere attains maximum size.  This unstable 2-sphere shrinks to a point at either vacuum ($\mu=0$ or $\pi$). There are two ways to cover the 3-sphere : (a) the Manton (or the finite mass) path (\ref{Manton2P}), (b) the constant mass (or the spherically symmetric) path (\ref{ours}) used in this paper. } \label{S3}
 \end{center}
\end{figure}

The $SU(2)$ gauge theory has Euclidean instanton solutions as the result of mapping spatial 3-sphere $S^3$ (in 4-dimensional Euclidean space) to the gauge manifold,  with homotopy $\pi_3\left( SU(2) \right) = \pi_3\left( S^3 \right) =\mathbb{Z} $.  That is, the Euclidean equations of motion (with $\hat t$ as the Euclidean time) have solitonic solutions where a spatial 3-sphere maps to the gauge field $A_i({\hat t}, r, \theta, \varphi)$.  In Minkowski (3+1) space in the electroweak theory, one considers static $A_i(\mu, r, \theta, \varphi)$ and  $\Phi(\mu, r, \theta, \varphi)$ where $0\le \mu < \pi$. (We choose $\left|n=0 \right \rangle =\left|\mu=0 \right\rangle$ as the reference vacuum.) At fixed $r$,  the point $p(\mu, \theta, \varphi)$ spans a 3-sphere, as shown in Fig. \ref{S3}. 
The $S^2$ swept by the usual polar angles $(\theta, \varphi)$ has zero size at $\mu=n \pi=0$ and $\mu=n\pi= \pi$, and maximum size at $\mu=n\pi=\pi/2$
corresponding to the sphaleron, an extremum (but unstable) solution of the static equations of motion. Time-dependence is then introduced when we treat $\mu(t)$ as a function of time.

A sphaleron is an unstable static solution of the classical equations of motion of the electroweak theory \cite{Manton1983,Klinkhamer1984} so its energy measures the height of the potential barrier that separates the $\left|n \right\rangle$ vacuum from the $\left|n+1\right\rangle$ vacuum. Here we are interested in the potential between two adjacent vacua, not just the height of the barrier. There are two approaches in finding this sphaleron potential, namely the Manton approach \cite{Manton1983} and the AKY approach\cite{Akiba1988}.
Let us review these two approaches and comment on their relation.


\subsection{The Sphaleron Potential}

It is convenient to work in the temporal gauge $A_0=0$ and then further impose the polar gauge $rA_r=0$ to fix all local gauge freedom. This implies that $A_r=0$ for all $r$ except may be at $r=0$. In fact, as we shall see, $A_r=-\mu \delta(r) \hat{r}\cdot{\vec \sigma}/2g$. Here, we can ignore this for the moment. In the spherically symmetric ansatz, we can write the fields in the following forms
\begin{align}
\label{ours}
   \tilde{\Phi} &= v\left(1-h(r) \right) U \begin{pmatrix}
   0\\ \cos \mu
   \end{pmatrix} + h(r) \begin{pmatrix}
   0\\ v
   \end{pmatrix} , \nonumber\\
   A_i &= {i\over g} (1-f(r)) U \partial_i  U^{\dagger} , \nonumber\\
 U &=\begin{pmatrix} 
      \cos \mu + i \sin \mu \cos{\theta}  &   - \sin \mu \sin{\theta} e^{i\varphi} \\
       \sin \mu \sin{\theta} e^{-i\varphi} &   \cos \mu - i \sin \mu \cos{\theta}
\end{pmatrix} , \nonumber\\
   \lim_{r\to 0}& {f(r) \over r}= h(0)=0, \quad f(\infty) = h(\infty) =1 ,
\end{align}
where $\{\mu,\theta,\varphi\}$ are the polar angles of $S^3$ and the $SU(2)$ matrix $U$ maps this $S^3$ to the group manifold $S^3_{SU(2)}$. With the above Higgs field profile $h(r)$ and gauge field profile $f(r)$, $\mu=0$ and $\mu={\pi \over2}$ correspond to the vacuum and the sphaleron respectively. Note that the asymptotic Higgs vacuum expectation value is always $\Phi^t=[0,v]$ for every $0 \le \mu < \pi$ (shown in Fig. \ref{figfar} (b)), which is different from the Manton setup \cite{Manton1983} (shown in Fig. \ref{figfar}(a)). This difference is explained in Sec. \ref{sec:comparison} as it is important in the determination of the mass. To avoid confusion, we shall call the above ansatz  (\ref{ours}) the constant mass construction.

The energy of the sphaleron path is given by 
\begin{align} \label{Vmu}
  V_M(\mu)=&{4\pi \over g^2} \int dr \left\{4f'^2+{8\over r^2}[f(f-1)]^2 \sin^2\mu +2r^2 h'^2+4m_W^2(f-h)^2 \right.   \nonumber\\
   & \left.+4m_W^2[f(h-1)(hf+f-2h)]\sin^2 \mu   + \frac{1}{2} m_H^2 r^2 (h^2 -1)^2 \sin^2 \mu \right\} \sin^2\mu ,
\end{align}
where the static equations of motion at the sphaleron ($\mu=\pi/2$) are
\begin{align}
\label{fhsol}
 r^2 f'' &= 2f(1-f)(1-2f) + m_W^2 r^2 h(f-1), \nonumber \\
 \left(r^2 h' \right)' &= 2h(1-f)^2 + \frac{1}{2} m_H^2 r^2 (h^2-1)h.
\end{align}
Using the above boundary conditions, $f(r)$ and $h(r)$ can be easily solved numerically (see Appendix A). As mentioned in Ref\cite{Klinkhamer1984}, the sphaleron solution indeed has topological number $\frac{1}{2}$. Its energy $E_{sph}=9.11$ TeV measures the potential barrier height.
As shown in Fig. \ref{S3}(b), varying $\mu$ from $0$ to $\pi$ spans a 3-sphere, which also goes from the $\left|n=0 \right\rangle$  vacuum over the potential barrier to the $\left|n=1 \right\rangle$ vauum. So it is a reasonable approximation to take the $f(r)$ and $h(r)$ solution of Eq.(\ref{fhsol}) for $\mu=\pi/2$ and insert them directly into $V_M(\mu)$ (\ref{Vmu}) to obtain $V(\mu)$ (\ref{mainresult}).


\section{The Effective Time-independent Schr\"{o}dinger Equation for the Chern-Simons Number}\label{sec:schro}

 In this section, we write down the effective one-dimensional time-independent Schr\"{o}dinger equation where the coordinate is the CS number $n$. Since the potential $V(\mu)$ (\ref{mainresult}) has already been evaluated, we only need to find the mass $m$. 
 
The Lagrangian (\ref{lagran}) has kinetic terms for both $A_{\mu}(x)$ and $\Phi (x)$ that include the time derivatives. Here, we are only interested in the change of the CS number $n$ (or $\mu$) as a function of time : we start with  the static solutions for $A_{i}(n, r, \theta, \varphi)$ and $\Phi (n, r, \theta, \varphi)$ and then introduce time-dependence only from the change in $n$, i.e., ${\dot n} =\frac{\partial n}{\partial t}$,  
$$ \frac{\partial A_{i}}{\partial t} =   \frac{\partial A_{i}}{\partial n} \frac{\partial n}{\partial t}, 
\quad \quad  \frac{\partial \Phi}{\partial t} =   \frac{\partial \Phi}{\partial n} \frac{\partial n}{\partial t}. 
$$ So the kinetic terms in the Lagrangian (\ref{lagran}) yield a kinetic term for $n$. That is, we set to determine the mass (which can be a function of $n$) to obtain the kinetic term $m(n) {\dot n}^2/2$. Once we have the Lagrangian for $n$, we quantize the system and write down the time-independent Schr\"{o}dinger Equation for $n$.

A comment is in order here.  In principle, we should allow explicit time dependence in both $A_{\mu}(x)$ and $\Phi (x)$ to obtain the kinetic term for $\mu$ in a fully gauge invariant approach. This requires allowing  $A_0$ to be non-zero.  $A_0$ then is determined implicitly by imposing Gauss law, the field equation obtained by varying with
respect to $A_0$ (or perhaps $A_{\mu}$). We can still impose other gauge conditions, for example, spherical symmetry together with a radial gauge condition $rA_r = 0$, so the mass density should be spherically symmetric. In practice, the role of $A_0$ is to subtract off from field variations the part that is just an infinitesimal gauge transformation. As a result, our estimate ignoring $A_0$ (or setting it to zero) should give an overestimate of the kinetic energy and hence the mass parameter $m$. Fortunately, as we shall see, this over-estimate will have little or no effect in the phenomenology discussed in this paper.

 It turns out that the variable $\mu$ used in Sec. 2 is a more suitable coordinate for our purpose. It is closely related to $n$ (this relation will be explained in Sec. 4),  
\begin{equation} 
\label{nmur}
n\pi=\mu -\frac{\sin(2\mu)}{2} ,
\end{equation}
which equals $\mu= n\pi$ at integer and half-integer values of $n$, i.e., solutions (extrema) of the static equations of motion.  
Let us promote $\mu$ to a time-dependent variable $\mu(t)$ and compute the Lagrangian in terms of $\{\mu, \dot{\mu}\}$
\begin{equation}
   L=\int d^3x {\cal L} ={ 1\over 2} m {\dot{\mu}^2\over (m_W)^2} - V(\mu),
\end{equation}
where the constant mass $m$ is given by 
\begin{align}
\label{ourmass}
   m={8\pi m_W\over g^2} \int d(m_Wr) \big[4(f(r)-1)^2  + 2 (m_Wr)^2(h(r)-1)^2 \big] 
    = 17.1 \, \mbox{TeV}.,
\end{align}
where $f(r)$ and $h(r)$ are the $\mu=\pi/2$ solutions obtained from Eq.(\ref{fhsol}).
Define a coordinate with length dimension $Q= {\mu/m_W}$ such that $L = {1\over2} m \dot{Q}^2 - V(Q)$ (to make the commutator dimensionless as well). From this Lagrangian, one can define the canonical conjugate momentum of $Q$, and hence the Hamiltonian $H$,
\begin{equation}
  \pi_Q= { \partial L \over \partial \dot{Q}} = m\dot{Q}, \quad  H = \pi \dot{Q} -L = {\pi_Q^2 \over 2m} + V(Q),
\end{equation}
By imposing quantization on the variable $Q$, $[Q,\pi_Q]= i $, $\pi_Q$ in $H$ can be replaced by $-i{\partial \over \partial Q}$, resulting in the following familiar one-dimensional time-independent Schr\"{o}dinger equation,
\begin{equation}   
   \left( - {1 \over 2m}{\partial ^2 \over \partial Q^2} + V(Q) \right) \Psi(Q) = E \Psi(Q),  \label{schro}
\end{equation}
where $\Psi(Q)$ is the eigen-wavefunction of $Q$ and energy $E$ is the eigenvalue. This equation is solved in Sec. \ref{sec:bloch}. Readers mostly interested in phenomenology may go directly there.  \\


\section{Comparison to the Original Manton Construction}\label{sec:comparison}

As pointed out earlier, the above evaluation of the $\mu$-independent mass $m$ is obtained after making two changes to the original Manton construction. Here we like to explain these two changes. Both are gauge changes so the potential $V(\mu)$ remains the same as that obtained by Manton. However, the static gauge transformations are no longer pure gauge changes when we attempt to extract the time-dependent kinetic term from the static solutions. The differences are illustrated in Fig. \ref{S3} and Fig. \ref{figfar} and summarized in the end of this section.

\subsection{Rendering the Mass Finite}

The original Manton construction starts with (following his notations) the following ansatz involving the unitary $SU(2)$ matrix $U^{\infty}$,

\begin{align}
\label{Mantoninfty}
   \Phi (\mu, r, \theta, \varphi) &= v[1-h(r)] \begin{pmatrix}
   0 \\ e^{-i\mu} \cos\mu   \end{pmatrix} + vh(r) U^{\infty} (\mu, \theta,\varphi) \begin{pmatrix} 0\\ 1 \end{pmatrix}, \nonumber\\
   A_i(\mu, r, \theta, \varphi) &= {i\over g} f(r) U^{\infty} \partial_i  (U^{\infty \dagger}) , \nonumber\\
 U^{\infty} &= \frac{1}{v}\begin{pmatrix}
 \Phi_2^{\infty *} &  \Phi_1^{\infty}  \\
 -\Phi_1^{\infty *} & \Phi_2^{\infty} \end{pmatrix} =
 \begin{pmatrix} 
    e^{i \mu}( \cos \mu - i \sin \mu \cos{\theta})  &    \sin \mu \sin{\theta} e^{i\varphi} \\
       -\sin \mu \sin{\theta} e^{-i\varphi} &  e^{-i\mu} (\cos \mu + i \sin \mu \cos{\theta})
\end{pmatrix} , \nonumber\\
 \Phi^{\infty}(\mu, \theta, \varphi) &= v \begin{pmatrix}
  \sin \mu \sin{\theta} e^{i\varphi} \\   e^{-i\mu} (\cos \mu + i \sin \mu \cos{\theta} ) \end{pmatrix} ,\nonumber\\
   \lim_{r\to 0}{f(r) \over r}&= h(0)=0, \quad f(\infty) = h(\infty) =1 .
\end{align}

\begin{figure}[h]
 \begin{center}
  \includegraphics[scale=.55]{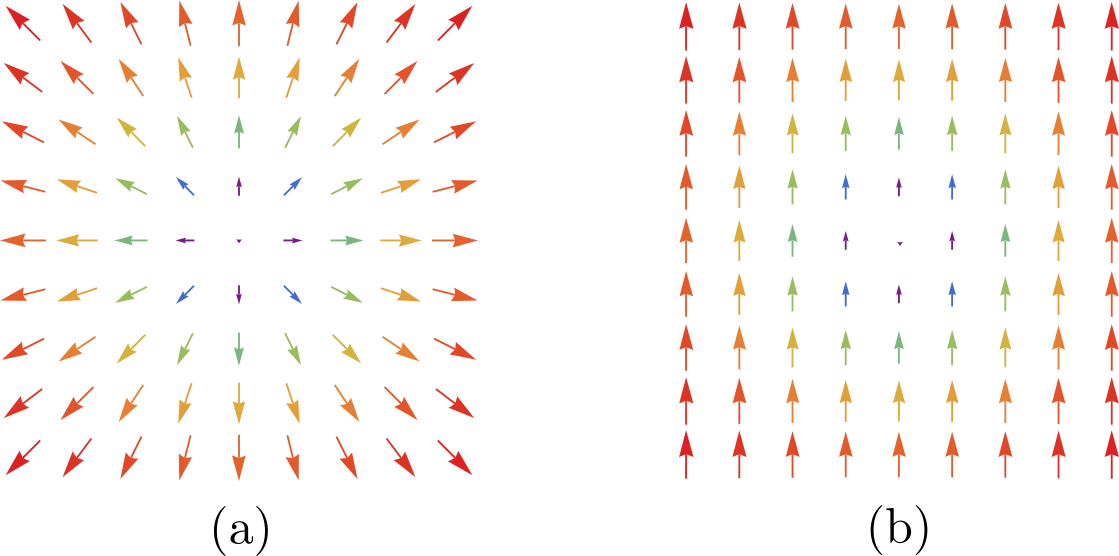}
  \caption{The direction of the vacuum expectation value of $\Phi$ at large distances. $[\re \Phi_1, \re \Phi_2]^T$ in (a) hedgehog gauge (\ref{Mantoninfty}), where $\Phi$ is rotated as $\mu$ varies at asymptotic distances, to reach a hedgehog shape at $\mu=\pi/2$; this introduces an infinite kinetic term; (b) unitary gauge at the sphaleron (\ref{ours}), where $\Phi$ always stays the same at large distances even as $\mu$ varies.} \label{figfar}
 \end{center}
\end{figure}

As $\mu$ ranges from  $0 \rightarrow \pi $, the Higgs field at $r \rightarrow \infty$ takes a hedgehog form, 
as illustrated in Fig. {\ref{figfar} a}. Note that this choice is not suitable for our purpose since the asymptotic Higgs field is in different gauges as $\mu$ goes from $0$ to ${\pi \over 2}$. It results in a divergent mass $m$.
This is because a change in $\mu$ results in a change in $\Phi$ at $r \rightarrow \infty$, even when the sphaleron is supposed to be localized. 
To remove this divergence, we simply make a gauge change to the fields in Eq.(\ref{Mantoninfty}) so at $r \rightarrow \infty$, $\Phi^t \rightarrow [0, v]$ independent of $\mu$. That is,

\begin{align}
\label{Mantonv2}
 {\hat \Phi} &= U^{\infty \dagger} \Phi = v(1-h)U^{\infty \dagger} \begin{pmatrix}
   0\\  e^{-i\mu} \cos \mu    \end{pmatrix} + v h \begin{pmatrix}
  0\\ 1   \end{pmatrix}, \nonumber\\
   {\hat A}_i &= U^{\infty \dagger} A_i U^{\infty} + \frac{i}{g} U^{\infty \dagger} \partial_i U^{\infty} = {i\over g} (1-f(r)) U^{\infty \dagger} \partial_i  U^{\infty}  \nonumber\\
 U^{\infty \dagger} &=   \begin{pmatrix}  e^{-i \mu}( \cos \mu + i \sin \mu \cos{\theta})  &    -\sin \mu \sin{\theta} e^{i\varphi} \\
       \sin \mu \sin{\theta} e^{-i\varphi} &  e^{i\mu} (\cos \mu - i \sin \mu \cos{\theta})
\end{pmatrix},  \nonumber\\
   \lim_{r\to 0}& {f(r) \over r}= h(0)=0, \quad f(\infty) = h(\infty) =1 .
\end{align}

This renders the mass $m$ finite so we refer to this as the finite mass construction. 

\subsection{Maintaining Spherical Symmetry}
 
Note the $e^{\pm i\mu}$ factor difference between Eq.(\ref{Mantonv2}) and Eq.(\ref{ours}). Let us write the unitary matrix $U^{\infty \dagger}$ in Eq.(\ref{Mantonv2}) in terms of the Pauli matrices and a 4-component vector ${\bf x}$,
$$U^{\infty \dagger}= x_4{\bf I} + i(x_i\sigma_i)$$ so that its determinant yields ${\bf x}\cdot{\bf x}=\sum x_i^2 =1$. We associate this unit vector ${\bf x}$ with a point $p(\mu, \theta, \varphi)$ in $S^3$, 
\begin{align}
\label{Manton2P}
  p (\mu, \theta, \varphi) 
   &=\begin{pmatrix}
    -\sin \mu \sin{\theta}\sin \varphi  \\ 
     -\sin \mu \sin{\theta} \cos \varphi \\ 
    - \sin \mu \cos \mu (1- \cos \theta ) \\ 
     \sin^2 \mu \cos \theta + \cos^2 \mu
\end{pmatrix}
\end{align}
This spans the $S^3$ as shown in Fig. \ref{S3}(a). 
On the other hand, $U$ in Eq.({\ref{ours}) yields
\begin{align}
\label{ours2}
  p (\mu, \theta, \varphi) &= \begin{pmatrix}
    -\sin \mu \sin{\theta}\sin \varphi  \\ 
    -\sin \mu \sin{\theta} \cos \varphi \\
     \sin \mu \cos \theta \\
     \cos \mu 
\end{pmatrix}
\end{align}
which spans the same $S^3$ but in a different way, as shown in Fig. \ref{S3}(b). 

As a point $p(\mu, \theta, \varphi)$ in $S^3$, each of the above two cases generates the same $S^3$ by generating $S^2$ for fixed $\mu$ and then varying $0 \leq \mu  \leq \pi$, but via two different paths.
The mass $m$ from the finite mass path (\ref{Mantonv2}) yields

\begin{align}
\label{Mmass}
   m(\mu) &= \frac{4\pi m_W}{g^2}\int d(\cos \theta) d(m_Wr) \bigg\{ \left[4+2(1-4\cos \theta +\cos^2\theta)\sin^2 \mu\right](f-1)^2  \nonumber\\
  &\quad\quad\quad\quad\quad\quad\quad\quad\quad\quad\quad  +\frac{1}{2}(m_Wr)^2\left[4+(1-\cos^2 \theta)\sin^2 2\mu\right](h-1)^2 \bigg\} , \nonumber \\
   &={8\pi m_W\over g^2} \int d(m_Wr) \left[4(f-1)^2 F(\mu) +2(m_Wr)^2(h-1)^2 G(\mu) \right],
\end{align}

where
\begin{equation}
 F(\mu) =1+ \frac{2}{3}\sin^2 \mu ,\quad G(\mu) = 1 + \frac{1}{6}\sin^2 2\mu,
\end{equation}
while our path (\ref{ours}) yields $m$ in Eq.(\ref{ourmass}).
We see in the mass integral (\ref{Mmass}) in the finite mass path (\ref{Mantonv2}) that the $\mu$-dependent terms in $m(\mu)$ are introduced by the  angular dependent terms, which breaks the spherical symmetry adopted in the original ansatz.  Observe in Fig. \ref{S3}(a) that, as $\mu$ varies, $p(\mu,\theta=0, \varphi)$ is fixed at $(0,0,0,1)^T$ while $p(\mu,\theta\neq 0)$ travels at different ``velocity" at
different $\theta$. This $\theta$-dependent motion in $S^3$ introduces
the $\theta$-dependence in the mass $m$ density, leading to a bigger mass $m(\mu)$ that depends on $\mu$.
To maintain the spherical symmetry, we instead choose $U$ in (\ref{ours}, \ref{ours2}); that is, we simply drop the $e^{\pm i\mu}$ factors in Eq.(\ref{Mantonv2}). This yields the constant mass $m$ (\ref{ourmass}) (i.e., with $F(\mu)=G(\mu)=1$ instead of the values given in Eq.(\ref{Mmass})). Note that the difference between these two cases can be treated as a pure gauge transformation, so the static potential $V(\mu)$ remains the same in each case. However, it is no longer a simple static gauge choice when we attempt to extract the time-dependent kinetic term from the static solution in the static gauge.


\subsection{Relation between the Chern-Simons Number and the Canonical Variable $\mu$ }

To discuss the topological baryon number, the Chern-Simons number is often used. However, one has to pay extra attention to it as it is not fully gauge invariant in a time slice and it is not necessarily equivalent to the topological baryon number. Writing the Chern character (\ref{chern}) in the Chern-Simons form,
\begin{align}
N = \frac{g^2}{16 \pi^2} \int d^4x \Tr\left[F_{\mu \nu} \tilde{F}^{\mu \nu} \right] = \int d^4x \partial_{\mu} K^{\mu},  \nonumber \\
K^{\mu}= \frac{g^2}{32\pi^2}\epsilon^{\mu\nu\rho\sigma}\left(F^a_{\nu\rho}A^a_{\sigma} -\frac{g}{3}\epsilon^{abc} A^a_{\nu}A^b_{\rho}A^c_{\sigma}\right).
\end{align}
So
\begin{equation}
\label{Kterm}
N = \int d^3x K^0 {\big |}_{t=t_0} + \int_{-\infty}^{t_0} \int_{S} \vec{K}\cdot d\vec{S}.
\end{equation}
We choose $\mu(t=-\infty)=0$ and $\mu(t=t_0)=\mu$.

Let us start with the fields in Eq.(\ref{Mantoninfty}), but dropping all the $\exp (\pm i \mu)$ factors, as we have just discussed. Since this ${\tilde A}_i$ (as well as the $A_i$ in Eq.(\ref{Mantoninfty})) drops off like $1/r$ asymptotically, the surface term in Eq.(\ref{Kterm}) does not vanish. 
One can make use of the residual $U(1)$ gauge transformation \cite{Witten1977a},
$$ \exp \left(- i \Omega (r) \hat{r}\cdot {\vec{\sigma}}\right), \quad \quad \Omega (r) = \mu \tanh(\beta r),$$ 
(where $\beta$ is large) to rotate ${\tilde A}_i$ to $A_i$ in Eq.(\ref{ours})
at large distances. Now, $A_i$ drops off exponentially so the surface term no longer contributes to $N$. This difference is related to the asymptotic direction of the Higgs vacuum expectation value (see Fig. \ref{figfar}). 
Such a transformation contributes a term  to the Chern-Simons number \cite{Jackiw2000}, so the 4-dimensional Chern number (\ref{chern}) becomes
\begin{align}
N= \int d^3x K^0 {\big |}_{t=t_0} + \frac{2 \Omega(r) -\sin(2\Omega(r))}{2\pi}{\Big |}^{r=\infty}_{r=0} = \frac{2\mu -\sin(2\mu)}{2\pi}
\end{align}
since $K^0$ vanishes. This is the relation (\ref{nmur}). To get to the $A_i$ at all $r$ as given in Eq.(\ref{ours}), we have to take $\beta \rightarrow \infty$. In this case, we find that $A_r=-\mu \delta(r) \hat{r}\cdot{\vec \sigma}/2g$ instead of $A_r=0$. (Note that the polar gauge condition $rA_r=0$ is still satisfied.) 
This $\delta (r)$ term has no impact in the evaluation of the potential $V(\mu)$ or the mass $m$.

Let us summarize in the diagram below the changes carried out in this paper. 

\begin{center}
\begin{tikzpicture}\label{flowchart}
     [align=center]
    \node(A1) {Manton (\ref{Mantoninfty})};
    \node[right = 5.5cm of A1]  (A2) {$\int\vec{K}\cdot d\vec{S} \neq 0$ };
    \node[right = 2cm of A2] (A3) {AKY  \cite{Akiba1988}};
    \node[below= 2cm of A1, text width =4cm] (B1) {Finite mass (\ref{Mantonv2}, \ref{Mmass}) (mass density violating spherically symmetry )};
    \node[below = 2.25cm of A2, text width =4cm] (B2) {Constant mass (\ref{ours}) \\ (spherically symmetric mass density)};
    \node[below = 2.2cm of A3] (B3) {Sec. \ref{MAKYm}};
    \draw [->] (A1) -- node[below] {remove $e^{\pm i\mu}$} node[above]{Fig. 2(a) $\to$ 2(b)} (A2) ;
    \draw [->] (A1) --node [left] {Fig. 3(a) $\to$ 3(b)} (B1);
    \draw [->] (B1) -- node[above]{Fig. 2(a) $\to$ 2(b)} (B2) ;
    \draw [->] (A2) -- node[right] {$\quad\;$ Fig. 3(a) $\to$ 3(b)} (B2);
    \draw [->] (A3) --(B3);
\end{tikzpicture}
\end{center}


\section{Comparison of the Manton and the AKY Constructions}
\label{MAKYm}

Here we compare the sphaleron potential obtained in the Manton approach versus that in the AKY approach. This comparison is possible only when both use a coordinate with a canonical kinetic term. We see that their masses differ by about 30 \% while the potentials are quite close.

\subsection{The AKY Potential for the Sphaleron}

The Manton construction of the path between a sphaleron and the vacuum comes from the picture of shrinking a $S^2$ loop in $S^3$ to a point. The study of this path is performed in a different manner by Akiba, Kikuchi and Yanagida (AKY) \cite{Akiba1988}, in which a static minimum-energy path is obtained by considering the most general spherically symmetric ansatz of the fields,

\begin{align}\label{AKY3}
  A_0^a &= \frac{1}{g} a_0(r,t)\hat{x}^a, \quad
  A_j^a = \frac{1}{g} \left[ a_1(r,t) \hat{x}_j\hat{x}_a + \frac{f_A(r,t)-1}{r}\epsilon_{jam}\hat{x}_m + \frac{f_B(r,t)}{r}( \delta_{ja} - \hat{x}_j\hat{x}_a) \right],  \nonumber \\
  \Phi & = \left( H(r,t) +  K(r,t)i \vec{\sigma}\cdot \hat{x} \right)\begin{pmatrix} 0\\ v
   \end{pmatrix}.
\end{align}
It is shown in \cite{Witten1977a,Ratra1988} that under these spherically symmetric field configurations, the $SU(2)$ gauge and Higgs sectors in the model (\ref{lagran}) reduce to a $(1+1)$-dimensional model,
\begin{align}
  \label{2dL}  \frac{g^2}{4\pi} {\cal L} =& -\frac{1}{4}r^2 f_{\mu\nu}f^{\mu\nu} + (D_{\mu} \chi)^{\dagger}D^{\mu} \chi -\frac{1}{2r^2}(|\chi|^2-1)^2  \nonumber \\
  & + 2m_W^2\left[ r^2 (D_{\mu}\phi)^{\dagger}D^{\mu}\phi  -  \frac{1}{2}(|\chi|^2 +1)|\phi|^2 + \re (\chi^* \phi^2) - \frac{m_H^2}{4} r^2 (|\phi|^2 -1)^2 \right],
\end{align}
where,
\begin{align}
 ds^2 = dt^2 -dr^2& , \quad f_{\mu\nu}= \partial_{\mu}a_{\nu} - \partial_{\nu}a_{\mu}, \quad \chi= f_A + if_B, \quad \phi = H + iK, \nonumber \\
& D_{\mu}\chi = (\partial_{\mu}-i a_{\mu} )\chi , \quad D_{\mu}\phi = (\partial_{\mu} -\frac{i}{2}a_{\mu}) \phi.
\end{align}

It preserves the $U(1)$ subgroup of $SU(2)$ via the gauge transformation $ a_{\mu} \rightarrow a_{\mu} +\partial_{\mu}\Omega, \; \chi\rightarrow \exp(i\Omega)\chi$ and $\phi \rightarrow \exp\left(\frac{i}{2}\Omega\right)\phi.$
In a time-independent setup, $a_0 = 0$ and $a_1=0$ are chosen to fix the local gauge. 
 However, static equations of motion from (\ref{2dL}) have no non-trivial solution when the winding number $N$ is not a half-integer, since they are not extrema of the energy functional.
The AKY static path is solved by minimizing the energy functional with respect to a fixed $N$ (\ref{chern}), namely, minimizing the functional,
\begin{equation} \label{lmultiplier}
   W[\chi,\phi] = E_{stat}[\chi,\phi] + \eta (N[\chi,\phi]-n),
\end{equation}
where $\eta$ is a Lagrange multiplier. The static equations of motion become 
\begin{align} \label{akyeom}
  \left[-\partial_r^2 +\frac{1}{r^2}(|\chi|^2-1)+m_W^2 |\phi|^2 \right] \chi = m_W^2\phi^2 - i m_W \zeta \partial_r\chi, \nonumber \\
  \left[ -\partial_r^2 +\frac{1}{2r^2}(|\chi|^2 +1) +\frac{1}{2}m_H^2 (|\phi|^2 -1) \right] (r\phi )= \frac{1}{r}\chi \phi^* ,
\end{align}
with $\zeta = {\alpha_W\eta}/{2\pi m_W}$. The asymptotic behavior of the solution is,
\begin{align}
\label{alpha1}
&r\rightarrow 0, \nonumber \\
 &\left\{ \begin{array}{l}
  \chi \approx e^{-i q} \left[1+c_A (m_Wr)^2 + i \frac{1}{3}(-c_Hc_K+c_A\zeta )(m_Wr)^3 \right] \\
  \phi \approx  e^{-i \frac{q}{2}} \left[ c_H +\frac{1}{12}c_H(c_H^2-1)(m_Hr)^2 +i c_k(m_Wr) \right]
  \end{array} , \right.  \nonumber\\
&r\rightarrow \infty, \nonumber \\
& \left\{ \begin{array}{l}
 \chi \approx  1+ \re(d_A e^{-\alpha r}) + i \re (d_B  e^{-\alpha r})   \\ 
  \phi \approx 1+ \frac{d_H}{m_Wr} e^{-m_Hr} -i \re \left(\frac{d_B}{\alpha^2 r^2}e^{-\alpha r} \right)
\end{array}  , \right. 
\end{align}
where
\begin{equation} \label{eqn:alpha}
  \alpha = \frac{m_W}{2}\left(\sqrt{4 -\zeta^2 } \pm i\zeta \right), \quad d_B = \frac{\zeta m_W \alpha d_A}{m_W^2- \alpha^2}.
\end{equation}
Note that a different boundary condition is chosen here compared to that in Ref\cite{Akiba1988}. These two sets of boundary conditions are equivalent up to a rigid $U(1)$ transformation of $\Omega=-q$, where $q$ is constant everywhere. Likewise, due to the aforementioned reason (see the summary diagram at the end of Sec. \ref{sec:comparison}), one has to keep the large distance vacuum in the same gauge in order to discuss the kinetic term.

Since the original AKY set of boundary conditions chosen in \cite{Akiba1988} is not in the unitary gauge, the surface term $\int \vec{K}\cdot d\vec{S}$ does not vanish;  one has to make use of the residual $U(1)$ gauge transformation with $\Omega(r\rightarrow\infty)\rightarrow -q$ exponentially and $\Omega(r=0)=0$ to make $A_i^a$ drop faster than $\frac{1}{r}$ such that $\int \vec{K}\cdot d\vec{S}=0$ \cite{Manton1983}. Such a transformation contributes a term $(q - \sin q)/2\pi$ to the CS number \cite{Jackiw2000}. The 4-dimensional Chern number (\ref{chern}) 
thus reduces to
\begin{align}
  N= \int d^3x K^0 +\frac{q-\sin q}{2\pi}  = \frac{1}{2\pi}\left( \int dr \re(i\chi^* \partial_r \chi) + f_B{\big |}^{r=\infty}_{r=0} \right) + \frac{q-\sin q}{2\pi},
\end{align}
where a reference vacuum of CS number $0$ is chosen at past infinity. We make a gauge transformation of $ \Omega(r)= -q \lim_{\beta \rightarrow \infty} \tanh(\beta r)$ to the AKY boundary conditions to obtain our boundary conditions in order to preserve regularity at the origin. This changes $a_1(r)$ in (\ref{AKY3}) to a $\delta$-function. This only changes the CS number and its effect is already explicitly included in the term $(q-\sin q)/2\pi$ while $\int d^3x K^0$ is safely calculated from the rest. 

The equations of motion (\ref{akyeom}) are solved numerically under the above boundary conditions and the AKY potential \cite{Akiba1988} is reproduced, as shown in Fig. \ref{massV}(a).


\subsection{Approximate Equivalence of the modified and the AKY Constructions}

Following the discussion in Sec. \ref{sec:schro}, we also examine the Lagrangian for AKY case. Promoting $n$ to a time-dependent function $n(t)$, the Lagrangian is written in $\{n,\dot{n}\}$,
  $L= \frac{1}{2}M(n) \dot{n}^2 - V(n)$,
where \footnote{$M(n)$ has dimension $[mass]^{-1}$ since $n$ is dimensionless.} 
\begin{align}
  M(n) = \frac{4\pi}{g^2}\int dr \bigg[ &(\partial_n f_A)^2 +(\partial_n f_B)^2   +2m_W^2 r^2 \left( (\partial_nH)^2 +(\partial_n K)^2 \right) \bigg].
\end{align}

\begin{figure}[h]
 \begin{center}
  \includegraphics[scale=.36]{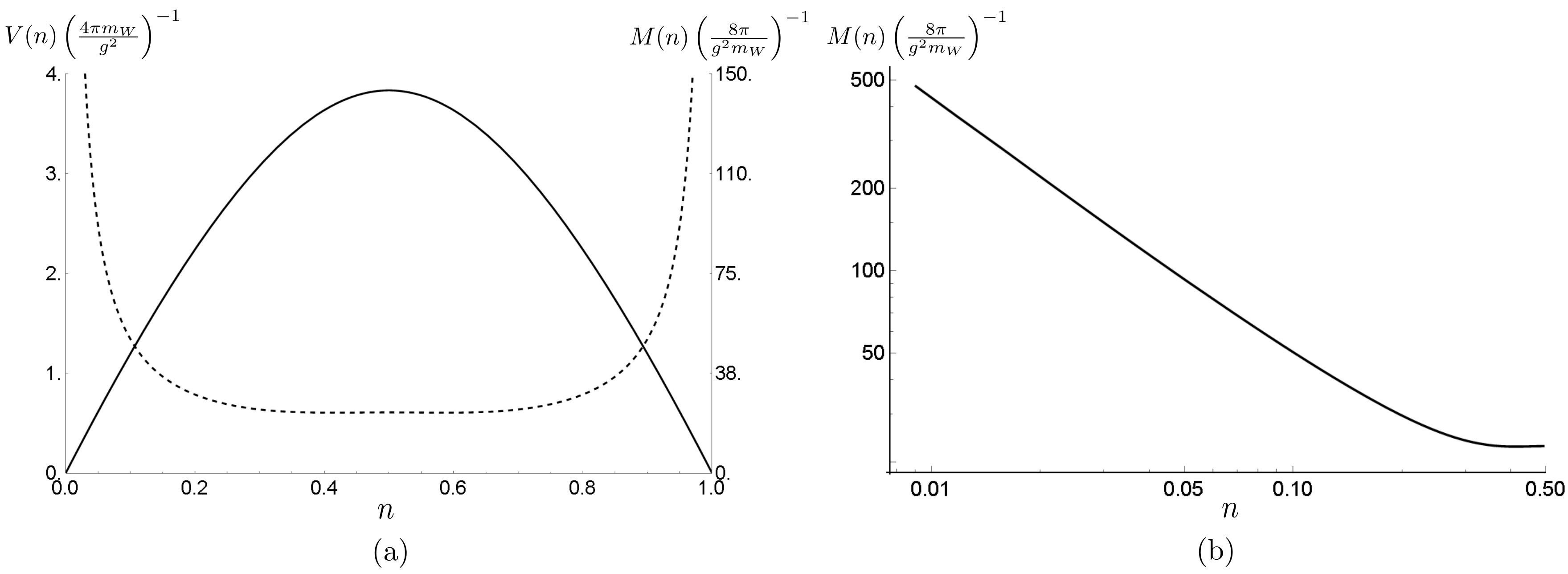}
  \caption{(a) Solid line: the AKY potential $V(n)$ is periodic in $n$ which has a cusp shape potential around each topological vacuum. Dashed curve: $M(n)$ is divergent at integer $n=0,1$. (b) $M(n)$ in log-log scale. It is approaching a straight line which gives $M(n)\propto n^{-0.98}$ as $n \rightarrow 0$. } \label{massV}
 \end{center}
\end{figure}

 Following the definition (\ref{lmultiplier}), one can verify by a Legendre transform that the potential $V(n)$ has a linear cusp near integer $n$ (see \cite{Akiba1988}),
\begin{equation}
 \left.\frac{d E(n)}{dn} \right|_{n=0} = \frac{n}{|n|} \frac{4\pi m_W}{\alpha_W} .
\end{equation}
It is interesting to see from Fig. \ref{massV} that $M(n)$ diverges at $n=0,1$.

This divergence is actually a nice feature which helps to remove the cusp of the AKY potential around a vacuum. The simple example illustrates how this happens : ${\dot y}^2/(4|y|) + |y|  \rightarrow {\dot x}^2 + x^2$ by the reparametrization  $y=x^2$. Both the divergence in the mass $1/(4|y|)$ and the cusp feature of the potential $|y|$ at $y=0$ vanish in the canonical variable $x$. If  $M(n)$ diverges as $n^{-1}$, AKY potential will become quadratic-like around vacuum which is similar to the Manton potential. 
We checked numerically the power of divergence in $M(n)$. It approaches $n^{-0.98}$ at $n=0.01$, instead of $n^{-1}$, though the latter value is possible as $n \to 0$. For phenomenological purposes, the exact limit of divergence power at very small $n$ does not affect our main result. We extrapolate the divergence power as $n^{-0.98}$ for small $n$ in discussions below. A similar behavior has been observed in an instanton calculation \cite{Bitar1978a}. 

\begin{figure}[h]
 \begin{center}
  \includegraphics[scale=.45]{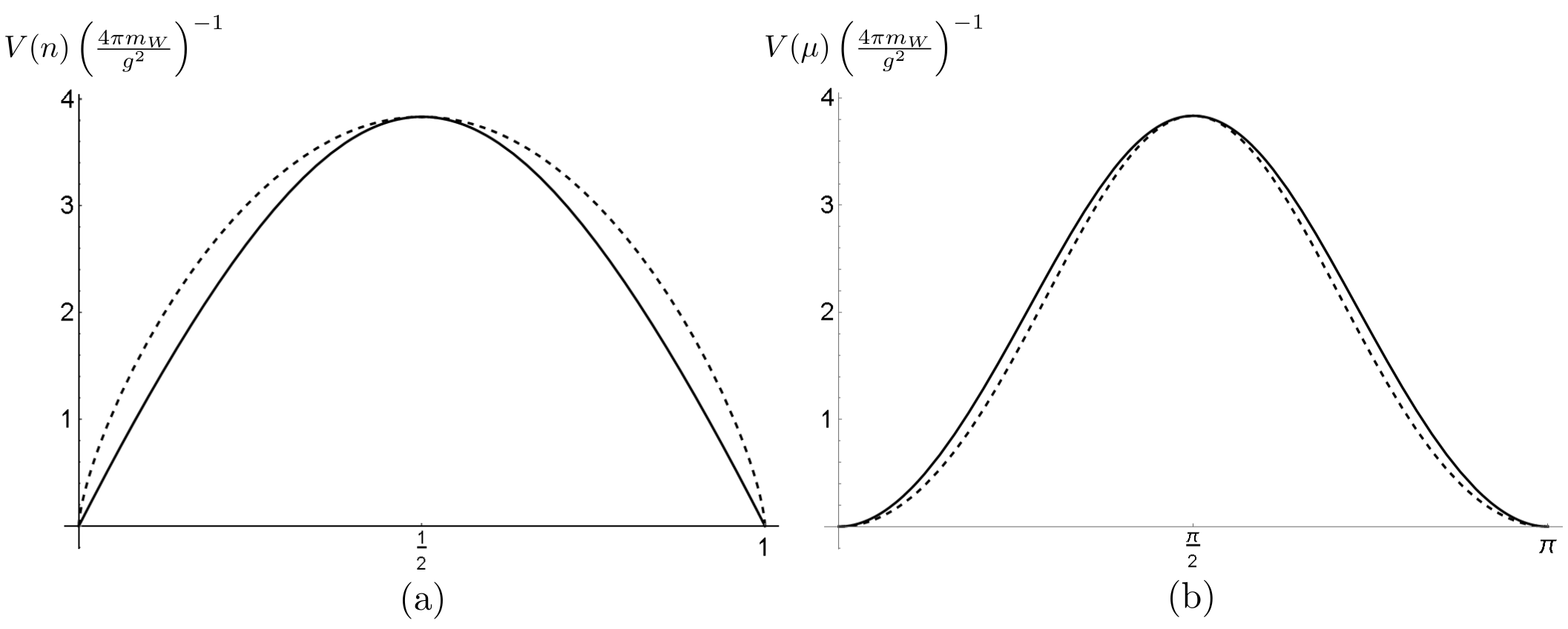}
  \caption{Comparison of AKY potential (solid line) and Manton potential (dashed line), (a) in CS number $n$, (b) in cannonical variable $\mu$. In the Manton case, $n=({2 \mu- \sin 2\mu})/(2\pi)$, which reduces to $n=\mu/\pi$ at half-interger values of $n$.}
   \label{akymanton}
 \end{center}
\end{figure}

After a reparametrization $ \mu =a \int \sqrt{M(n)}dn $, where $a$ is fixed by setting $\mu =\frac{\pi}{2}$ at the sphaleron, the Lagrangian with a canonical kinetic term is obtained,
$ L = m {\dot{\mu}^2}/(2m_W^2) - V(\mu).$
Numerically, $m\simeq 22.5$ TeV which is around 1.3 times the mass $m=17.1$ TeV in (\ref{mainresult}).  We observe that the shape of AKY potential $V(\mu)$ displayed in Fig. \ref{akymanton} is now quite close to the Manton potential  in this variable.

We again quantize the variable $Q={\mu}/{m_W}$ and solve for the eigenvalues of the Hamiltonian to get the band structure for the AKY case. Results are presented in Sec. \ref{sec:bloch}. Based on our analysis, we see that $\mu/\pi$ (instead of $n$) is the best choice as the CS number.


\section{Bloch Wave and Band Structure}\label{sec:bloch}

\subsection{Solving the Schr\"{o}dinger Equation}

In this section we solve the Schr\"{o}dinger equation (\ref{schro}) in both the Manton and the AKY construction. Since $V(Q)$ is periodic, the energy eigen-function of $H\psi = E\psi$ should be Bloch-waves, \footnote{Similar to the $\theta$-vacuum in QCD, the Hamiltonian eigenstates should be superpositions of all local states, $\left| k\right\rangle = \sum_n e^{i n k\pi/m_W } \left| n\right\rangle$.}
\begin{equation}
  \left\langle Q | k\right\rangle= \psi_k(Q) = e^{ik Q}u_k(Q), \;  u_k(Q)=u_k\left(Q+ {\pi \over m_W} \right).
\end{equation}
It is well known that the energy spectrum of a periodic potential has a band structure: continuous bands (solutions to the Schr\"{o}dinger equation (\ref{schro})) of certain widths separated by bandgaps (i.e., regions with no solution). 

It is standard to call the edge energies of the bands as eigenvalues and the corresponding wave functions as eigenfunctions. Note that $V(Q)$ is symmetric with respect to both $Q=0$, i.e., $V(Q)=V(-Q)$, and $Q=\pi/2m_W$, i.e., $V(Q+\pi/2m_W)=V(-Q+\pi/2m_W)$; so there are 4 types of eigenfunctions, namely $(SS)$, $(SA)$, $(AS)$ and $(AA)$, where the first letter denotes symmetric $(S)$ or anti-symmetric $(A)$ about $Q=0$ and the second letter denotes that about $Q=\pi/2m_W$. $(SS)$ and $(AA)$ eigenfunctions have period $\pi/m_W$ while $(SA)$ and $(AS)$ eigenfunctions have period $2\pi/m_W$. Starting from the (SS) ground state (i.e., lower edge energy of the first allowed band) as the energy is increased, the eigenfunctions have period $ [1, 2, 2, 1, 1, 2, 2, 1, 1, .....]\pi/m_W$.
One can approximate the band edge energies by the semi-classical method \cite{Sukhatme1999},
\begin{align}
\label{WKBS}
 \int_0^{Q_0}p(Q,E) dQ  = K {\pi \over 2} \pm \arctan\left[\tanh\left(\int_{Q_0}^{\pi\over 2m_W}p(Q,E)dQ\right)\right], \; K \in \mathbb{N}
\end{align}
where $p(Q,E) = \sqrt{2m|E- V(Q)|}$ and $Q_0$ denotes the classical turning point in the interval $[0, \pi/2m_W]$. 
The lowest band has edge energies with $K=0$ $(SS)$ and $K=1$ $(AS)$, the next band has edge energies with $K=1$ $(SA)$ and $K=2$ $(AA)$, and so on. This approximation is very accurate when the integral inside $\tanh$ in (\ref{WKBS}) is large.

It is convenient to solve the Schr\"{o}dinger equation (\ref{schro}) in momentum space using Bloch-waves,
\begin{equation}\label{eqn:kspacebloch}
  \sum_{l=-\infty}^{\infty}\left[ {1\over 2m}(2l+k)^2 \delta_{l,m} + V_{l-m} \right]u_{k,l} = E\,u_{k,m},
\end{equation}
where $V_{l-m},\; u_{k,m}$ are the Fourier coefficients of $V(Q),\; u_k(Q)$ respectively.
Here we solve Eq.(\ref{eqn:kspacebloch}) numerically for the band structure. Table \ref{tab:bandwidth} lists the band center energies and their bandwidths near the top of the potential barrier and around the bottom of the potential $V(Q)$. Six bands near the top are shown in Fig. \ref{pic:band}.  In Fig. \ref{widths2}, we  show that the logarithms of the bandwidths follows a linear curve. Since the band gap sizes change relatively slowly, it is easy to obtain the approximate band energies and their bandwidths for those not shown in Table \ref{tab:bandwidth}.  

\begin{figure}
\begin{center}
  \includegraphics[scale=0.9]{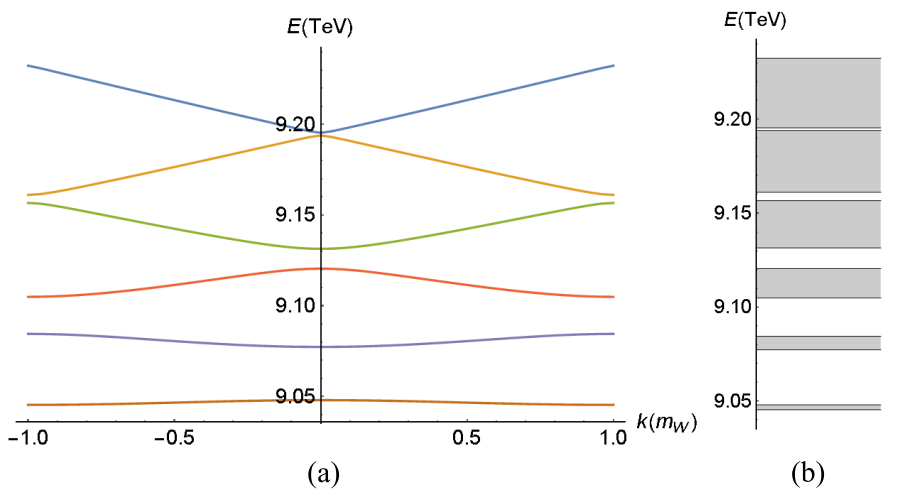}
   \caption{The (reduced) Brillouin zone plot (6 conducting bands shown) near the barrier top for the Manton case:  (a) The energy eigenvalue $E(k)$ of the Schr\"{o}dinger equation (\ref{schro}) is shown as a function of the propagation wave vector $k$ (in units of $m_W$); (b) The dark regions are the conducting bands while the gaps are regions where there is no solution to the Schr\"{o}dinger equation ((\ref{schro}) or (\ref{eqn:kspacebloch})). The energy spectrum is almost continuous above the barrier while bandwidths below the barrier decrease exponentially as the energy goes down. The band structures in the AKY case are similar (see Table \ref{tab:bandwidth}).}\label{pic:band}
\end{center}
\end{figure}

\begin{table}
\begin{center}
\begin{tabular}{cc|cc}
 \multicolumn{2}{c|}{Manton}  &\multicolumn{2}{|c}{AKY} \\ \hline
 Band Center & Width    & Band Center & Width  \\ 
 Energy(TeV) &    (TeV)        & Energy(TeV) & (TeV)   \\     \hline
 9.113 & 0.01555  &  9.110 & 0.01134  \\
 9.081 & 7.192$\times 10^{-3}$ & 9.084 & 4.957$\times 10^{-3}$ \\
 9.047 & 2.621$\times 10^{-3}$ & 9.056 & 1.718$\times 10^{-3}$\\
 9.010 & 8.255$\times10^{-4} $ & 9.026 & 5.186$\times 10^{-4}$\\
 8.971 & 2.382$\times 10^{-4}$ & 8.994 & 1.438$\times 10^{-4}$ \\
 8.931 & 6.460$\times 10^{-5}$ & 8.961 & 3.747$\times 10^{-5} $\\
 8.890 & 1.666$\times 10^{-5}$ & 8.927 & 9.279$\times 10^{-6}$\\
 8.847 & 4.114$\times 10^{-6}$ & 8.892 & 2.198$\times 10^{-6}$ \\
 8.804 & 9.779$\times 10^{-7}$ & 8.857 & 5.008$\times 10^{-7}$\\
 8.759 & 2.245$\times10^{-7}$ & 8.802 & 1.101$\times 10^{-7} $\\
 8.714 & 4.993$\times 10^{-8}$ & 8.783 & 2.341$\times 10^{-8}$\\
 8.668 & 1.078$\times 10^{-8}$ & 8.745 & 4.828$\times 10^{-9}$\\
 8.621 & 2.262$\times 10^{-9}$ & 8.707 & 9.673$\times 10^{-10}$\\
 8.574 & 4.622$\times 10^{-10}$ & 8.668 & 1.886$\times 10^{-10}$ \\
 8.526 & 9.210$\times10^{-11}$ & 8.628 & 3.580$\times 10^{-11}$ \\
 8.477 & 1.792$\times 10^{-11}$ & 8.588 & 6.622$\times 10^{-12}$\\
 8.428 & 3.411$\times 10^{-12}$ & 8.548 & 1.211$\times 10^{-12}$\\
 8.379 & 6.395$\times 10^{-13}$ & 8.506 & 2.167$\times 10^{-13}$\\
 8.328 & 1.208$\times 10^{-13}$ & 8.465 & 3.553$\times 10^{-14}$ \\
  $\vdots $  &   $ \vdots$   &    $\vdots $  &       $\vdots$   \\
   0.3084  & $\sim 10^{-169}$  & 0.3146 &$ \sim 10^{-204}$ \\
   0.2398  & $\sim 10^{-171}$  & 0.2454 & $\sim 10^{-207}$ \\
   0.1712  & $\sim 10^{-174}$  & 0.1759 & $\sim 10^{-209}$ \\
   0.1027  & $\sim 10^{-177}$  & 0.1061 & $\sim 10^{-212}$ \\
   0.03421 & $\sim 10^{-180}$  & 0.03574 & $\sim 10^{-216}$ \\
\end{tabular}
\end{center}
 \caption{Some of the top and the bottom band energies and their widths (in TeVs) are shown. There are 148 bands up to $E_{sph}= 9.11$ TeV in the Manton case and 164 bands in the AKY case. The band gap is about 70 GeV at low energies and decreases to about 30 GeV close to $E_{sph}$.}\label{tab:bandwidth}
\end{table}

\begin{figure}
 \begin{center}
  \includegraphics[scale=.6]{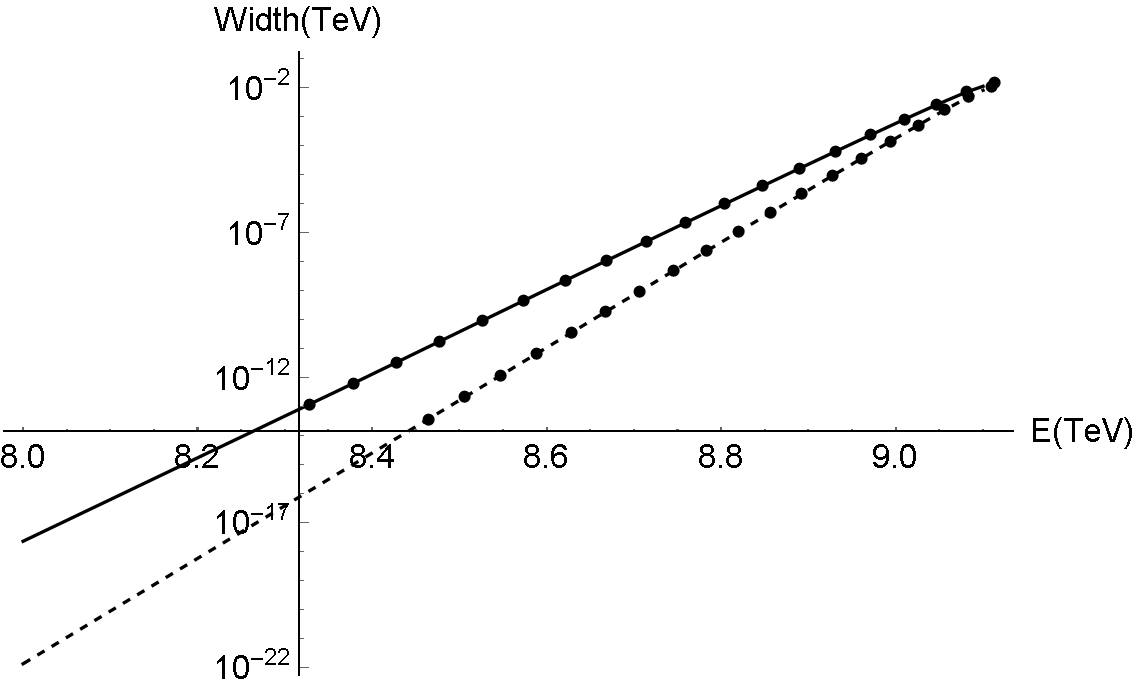}
  \caption{Bandwidths of Manton path (solid line) and AKY path (dashed line) decrease exponentially as energy decreases. This quantitative behavior obtained numerically fits well with that obtained in the WKB approximation (\ref{WKBS}).}  \label{widths2}
 \end{center}
\end{figure}

In the semi-classical approximation, a typical wavefunction with energy $E$ will have an energy spread $\delta E$ that is bigger than the width of band that it overlaps with. So it will have a tunneling probability amplitude that goes like
$$\Delta (E) \simeq \frac{\rm band width}{{\rm band gap}+{\rm band width}}$$ 
where the bandgap is around 70 GeV towards the bottom of the potential. With an exponentially small bandwidth for low energies, the tunneling probability is clearly exponentially suppressed, as expected. 
As the energy $E$ approaches $E_{sph}$ from below, the band widths increase quickly so the exponentially suppressing tunneling factor begins to vanish, so $\Delta (E) \rightarrow 1$. As we go above $E_{sph}$, the band gaps shrink as well so essentially there is no tunneling suppression. This expected property is shown in Fig. \ref{pic:band} and Fig. \ref{pic:rubakov}. In fact, for any given $k$ away from the edges, the wavefunction $\psi_k(Q)$ for $E(k) > E_{sph}$ is well approximated by a single free plane wave mode, i.e., we have the time-dependent wavefunction,
$$\psi_k(Q, t) \sim \exp (ipQ -i{\cal E}t)$$
where the "momentum" $p$ and the effective energy $\cal E$ depend on $E(k)$ and the average of the potential $<V>= 4.197$ TeV,
  $${\cal E} = E(k) - <V>  = \frac{p^2}{2m}$$
Note that the mean velocity is given by $<v>= p/m$, where $|p| > 13$ TeV for $E(k) > E_{sph}$.

Note that the bandwidths $\Delta$ for bands close to the bottom are well approximated by tight-binding model and the WKB method,
\begin{equation}
  \Delta = \left\langle n\pm 1|H|n\right\rangle \sim \exp\left( -\frac{2\pi}{\alpha_W} c \right),
\end{equation} 
where $c$ is a ${\cal O}(1)$ factor. Note that $c=1$ for pure gauge theory, and $c\sim 2.2$ and $c\sim 2.7$  respectively in the above two evaluations in Table \ref{tab:bandwidth}. This can be understood as the size of the sphaleron is fixed by the  electroweak breaking scale. Following Ref\cite{Hooft1976a}, we have
\begin{equation}
\label{tHooft3}
\Delta \sim  \exp[-S_A-S_H] = \exp \left[-\frac{2\pi}{\alpha_W}-\frac{\pi}{\alpha_W} (m_W \rho)^2\right],
\end{equation}
where $\rho$ is the sphaleron size, which is of order $1/m_W$. This gives a value not far from the tight-binding result, where the band width measures the tunneling amplitude. It is not surprising that AKY has a bigger value for $c$, as its sphaleron size is in general bigger than that in the Manton case.  In pure gauge theory, the instanton size $\rho$ is integrated over so the second term in the exponent in Eq.(\ref{tHooft3}) drops out.

At very low two-body scattering energy $E_2$ (with some energy spread), we see that the probability of any ($B+L$)-violating  process is exponentially suppressed. As we increase $E_2$, such processes will be exponentially less suppressed. This behavior is shown by the solid curve in Fig. \ref{pic:rubakov}. The two pass band structures in Table 1 yield almost the same curve (which is really two curves overlapping with each other). As the energy passes $E_{sph}$, the exponential suppression factor vanishes. 

\begin{figure}[h]
 \begin{center}
  \includegraphics[scale=.5]{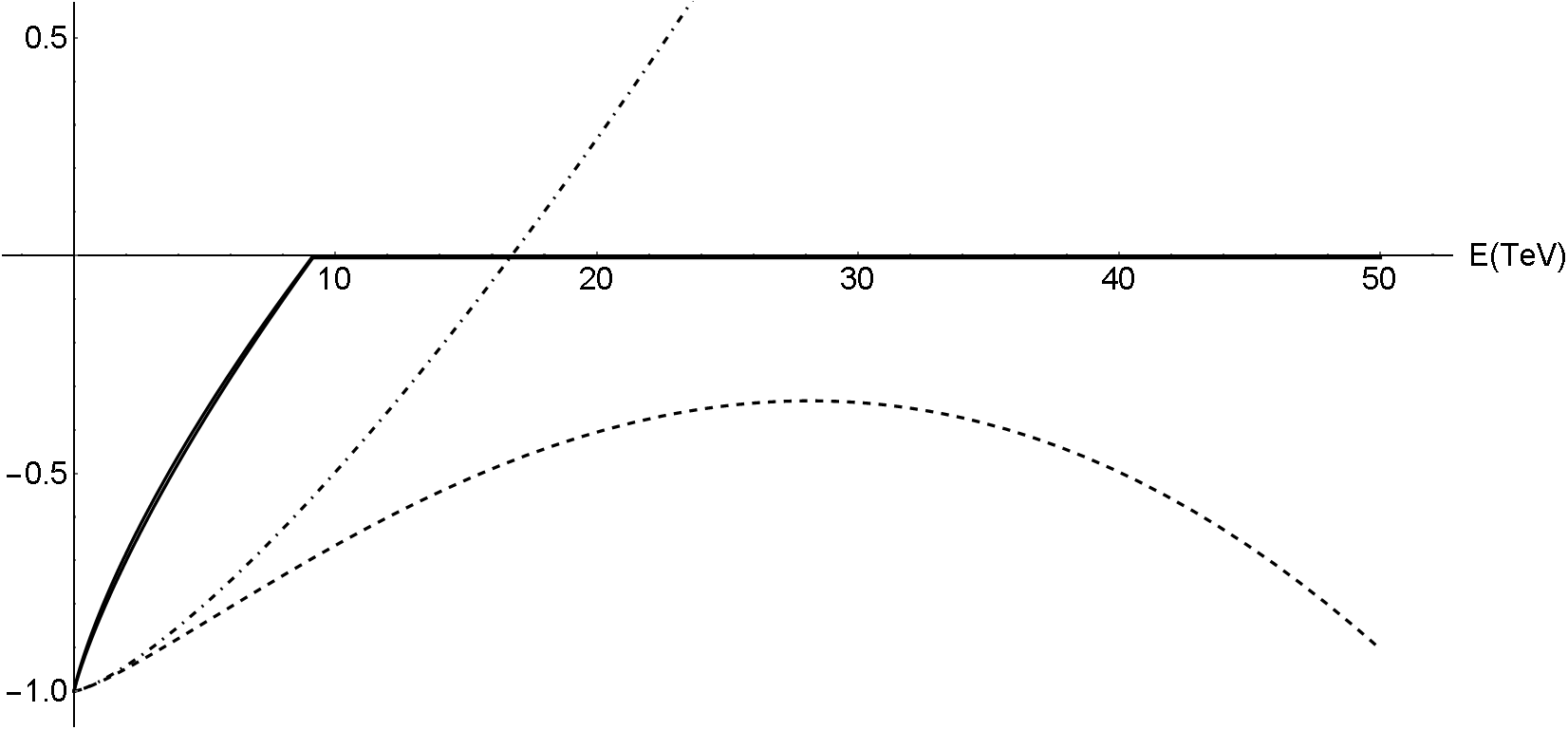}
  \caption{Comparison between the exponential suppression factor of (B+L)-violating processes in our case (solid line) and previous estimates. For ease of visualization, we normalize all cases to $-1$ at zero energy. The curves from the two band structures in Table 1 actually overlap to form a single solid curve that takes zero value above 9 TeV. The exponent factor (inside the square bracket) in Eq.(\ref{Espinosa}) is plotted here. The first order (dot-dashed) and second order (dashed) approximation in Eq.(\ref{Espinosa}) diverges differently at large $E_2$. } \label{pic:rubakov}
 \end{center}
\end{figure}


\subsection{Comparing with Earlier Studies}

Earlier works are also presented in Fig. \ref{pic:rubakov} for comparison. As we increase $E_2$, it is generally believed that such processes will be exponentially less suppressed. This behavior has shown up in perturbative calculations, and may be summarized in Fig. \ref{pic:rubakov}, where, for ease of comparison, we normalize all the curves for the various rates to the same value at $E_2 \simeq 0$. They have been captured in the following formula \cite{Espinosa1990,Ringwald1990a,Zakharov1992,Khlebnikov1991b,Porrati1990,Khoze1991},
\begin{align}
\label{Espinosa}
\sigma(\Delta n=\pm 1) \sim \exp \bigg( c \frac{4\pi}{\alpha_W}\bigg[ -1 +\frac{9}{8} \left(\frac{E_2}{E_0}\right)^{4/3}   - \frac{9}{16}\left(\frac{E_2}{E_0}\right)^{2} + . . .\bigg] \bigg),
\end{align}
where $E_0 = \sqrt{6} \pi m_W/\alpha_W =$ 15 TeV and $c \sim 2$.
The perturbative analysis of $|\Delta n| =1$ processes was started 
in \cite{Espinosa1990,Ringwald1990a}, which show that the inclusive cross-section rises exponentially with energy. The authors pointed out that unitarity breaks down if one extrapolates the result to high energies. The $E_2^{4/3}$ term in the exponent factor (in square brackets) in Eq.(\ref{Espinosa}) was later obtained \cite{Zakharov1992,Khlebnikov1991b,Porrati1990,Khoze1991,Mueller1991}; this is captured by the dot-dash curve in Fig. \ref{pic:rubakov}, which includes only the $E_2^{4/3}$ term (and the $-1$ term) in the exponent factor (in square brackets) in Eq.(\ref{Espinosa}).
This clearly violates unitarity, implying that its validity is at best limited. Including the next $E_2^2$ term in the exponent in Eq.(\ref{Espinosa}), its behavior is shown by the dashed curve in Fig. \ref{pic:rubakov}. Additional terms with higher power dependence on $E_2$ are known to exist. So this approach yields an inconclusive result.  
Non-perturbative studies have also been carried out  in toy models (see e.g., Ref\cite{Son1994}). Such analyses seem to imply that the rate of ($B+L$)-violating processes is always exponentially suppressed to a level at which they will never be observed in the laboratory at any finite energies \cite{Rubakov1996}. 

Expressed in terms of the WKB formula  for tunneling through a potential barrier $V(Q)$ (with maximum value $V_{\rm max} \sim m_W/\alpha_W$) at energy $E \sim 0$, one has a tunneling amplitude $\sim \exp (-\int \sqrt{2mV} dQ)$. Comparing this to the tunneling amplitude $\sim \exp (-2\pi/\alpha_W)$ estimated in earlier works in the semi-classical approximation, we find that mass $m \sim m_W/\alpha_W$. That is, the existence of the mass $m$ and its order of magnitude value have been implicitly assumed already in earlier studies.  We simply bring this semi-classical study to a fully first quantized analysis. Our approach includes the kinetic term for the CS number $\mu/\pi$, with $Q=\mu/m_W$ playing the role of the spatial coordinate. This allows us to carry out a fully first quantized analysis that incorporates the periodicity property of the sphaleron potential. The quantized version exhibits the important resonant tunneling effect via the Bloch theorem, with a simple band structure, where tunneling inside a pass band is unsuppressed. This crucial feature is absent in the earlier studies.


\section{Phenomenology}\label{pheno}

All the ($B+L$)-violating processes conserve electric charge, QCD color and the ($B-L$) number. The description of ($B+L$)-violating processes is cleanest  in $e^+e^-$ collisions, so we start with this. However, the energy needed will not be reached in the foreseeable future. So we turn to proton-proton collisions, which offer a much better chance of detection. This is particularly interesting at the Large Hadron Collider (LHC).

\subsection{$e^+e^-$ Collisions}

Let us first consider the following $\Delta(B+L)=6$ process in an $e^+ e^-$ annihilation starting in the vacuum state $\left|n=0\right\rangle$ going to $\left|n= + 1 \right\rangle$,
\begin{align}
e^+ + e^- \to  l_e + l_{\mu} + l_{\tau} + q_{(1)}+q_{(1)}+q_{(1)} +q_{(2)}+q_{(2)}+q_{(2)}+q_{(3)}+q_{(3)} +q_{(3)}  + X,
\end{align} 
where the subscripts are family indices. Here, 12 fermions are produced, a lepton and 3 quarks from each family. This and all scatterings discussed in this paper are meant to be inclusive processes; that is, they may include (as denoted by $X$) any number of $W^{\pm}$, Z and Higgs bosons, mesons and photons as well as  fermion-anti-fermion pairs, where $X$ has zero net baryon number and zero net lepton numbers but with an electric charge to preserve charge conservation of the process.  With 3 b-quarks from the 3rd family, the 12 fermions require about 20 GeV of energy, so we do not expect any phase space constraint in the ($B+L$)-violating processes. We do not address the issue of the number distribution of the massive bosons in this study. 
Instead of 3 quarks from each family, all quarks from the same family can also be produced, but such processes will be CKM mixing angle suppressed.  Equally likely is the $\Delta n=-1$ process.

As an illustration, suppose we have $e^+e^-$ collisions at about a few hundred GeV, and we can tune the energy over a wide range. If the energy of the $e^+e^-$ pair does not cover a pass band, then the above ($B+L$) violating processes will not happen. Let us assume that the $e^+e^-$ pair energy hits a pass band at energy $E \sim 100$ GeV (the second band in Table 1), which has a width of (using the more optimistic value in Table 1) $10^{-177}$ TeV. Taking the energy spread of the electron beam to be $100$ MeV, we see that only an exponentially small fraction (i.e., $10^{-173}$) of the $e^+e^-$ pair will lie inside the band.  So such ($B+L$)-violating processes are indeed exponentially suppressed.

The ($B+L$)-violating process in $e^+e^-$ collisions is very much suppressed due to the narrow band widths unless one can reach energies close to but below the sphaleron energy of 9 TeV. If one can reach that energy, say a couple of hundred GeVs below $E_{sph}$ and has a narrow enough beam energy spread of about 1 GeV, one can tune the $e^+e^-$ energy to sweep over one band at a time to enhance the signal-to-background ratio.

Unfortunately, a 9 TeV $e^+e^-$ (or $\mu^+\mu^-$) collider is not feasible in the foreseeable future; so let us turn to proton-proton colliders, which offer a better chance of detection in the near future.
   

\subsection{Proton-Proton Collisions}

Here, we like to explore ($B+L$)-violating processes in proton-proton collisions based on the above Bloch wave analysis. In pp collisions, the fundamental processes are quark and/or gluon scatterings. As a quark's momentum inside a proton has a wide parton distribution, the fundamental parton-parton center of mass energy has a very wide spread relative to a typical band width. To enhance the cross-sections, let us consider the case when both partons are valence quarks.

As an example, consider the following $\Delta n=-1$ process, 
\begin{equation}
u_{1L}+ u_{2L} \rightarrow e^+\mu^+ \tau^+ {\bar b}_1{\bar b_2} {\bar b}_3{\bar c}_1{\bar c}_2{\bar s}_3{\bar d}_3+ X,
\end{equation} 
where the subscripts are the color indices and $X$ stands for all other particles. In this inclusive scattering, the spectator quarks from the two protons, namely $u_2d_3$ and $u_1d_3$, can combine with 4 anti-quarks to form 4 mesons, leaving behind a single anti-baryon. 

For $\Delta n=+1$ process (\ref{qqBL+1}), consider, for example, 
\begin{equation}
u_{1L}+ u_{2L} \rightarrow e^-\mu^- \tau^-b_1b_2t_3c_1c_2c_3u_1u_2u_3d_1u_2 + X,
\end{equation}
a single pp collision can produce 5 baryons (plus baryon-anti-baryon pairs). Similarly, one can consider quark-gluon and gluon-gluon scatterings. Since gluons have no weak interactions, they must be converted to quarks or produce quarks for the ($B+L$)-violating processes to take place. Similar behavior can take place in an electron-proton collider.
All these scatterings typically will produce additional massive bosons: Higgs Bosons, $W^{\pm}$ and $Z$ Bosons, photons as well as mesons.


\subsection{LHC Physics}

Let us consider LHC run at 8 TeV. The center-of-mass energy for any pair of quarks has energy $E_c < 8$ TeV. For any of the above processes to be unsuppressed by an exponential tunneling suppression factor, it has to be inside a conducting band. For such a band with energy $E$ less than 8 TeV,  the band width $\Delta E \ll 10^{-20}$ TeV. So at most a tiny fraction (of order $10^{-20}$) of the quark-quark collisions take place within such a band. Even when collision events cover many bands, we do not expect any observable signal, as the band width decreases exponentially fast as the band energy decreases. So not a single ($B+L$)-violating event is expected.

Now the situation can be very different in the LHC runs at 13 and 14 TeV. In the actual electroweak theory with $\sin^2 \theta_W =0.23$, the estimate is $E_{sph} =$ 9.0 TeV. Here, the $qq$ scatterings can reach  $E (qq) >  E_{sph} =$ 9.0 TeV. A typical process can happen at energies covering a few dozen bands just below $E_{sph}$, where the band width varies from 10 GeV to MeVs. Therefore, there is a reasonable fraction of the scatterings taking place inside some bands, so ($B+L$)-violating events have a chance to be observed. Let us make a rough estimate of the event rate.

Among parton-parton scatterings, only left-handed quarks have direct electroweak interactions. 
Let $f(E(pp),E_c(q_Lq_L))$ be the fraction of $q_Lq_L$ scattering with energy $E(q_Lq_L)>E_c(q_Lq_L)$ at proton-proton collision energy $E(pp)$. 
Since the band width drops exponentially fast as $E(q_Lq_L)$ decreases, we may choose to introduce (rather arbitrarily) $E_c(q_Lq_L) \sim 9.0$ TeV to simplify the estimates.  Phenomenologically, with energies in units of TeVs, $f(14, 9)\sim10^{-6}$ to $10^{-8}$, based on the known parton distribution function for left-handed valence quarks \cite{Martin1998,Pumplin2002,Martin2009}. 
Among other parton-parton scatterings such as gluon-gluon,  gluon-quark, $q_Rq_R$ or $q_Lq_R$ scatterings, the partons have to convert to electroweak interacting particles first, which suppress their contribution to the ($B+L$)-violating process. The $q{\bar q}$ scatterings are expected to contribute little as well. Even in $q_Lq_L$ scatterings at energy above $E_c(q_Lq_L)$, radiation of a hard gluon can lower the energy of the $q_Lq_L$ pair to below $E_c(q_Lq_L)$. So we have to exclude the QCD $q_Lq_L$ cross section as well. Let the fraction of $q_Lq_L$ events that can participate in ($B+L$)-violating processes be $F_{EW} \sim \sigma_{EW}(q_Lq_L)/\sigma_{QCD}(q_Lq_L)$.  We expect $F_{EW} $ to be relatively small.

Even when a $q_Lq_L$ scattering can go through a ($B+L$)-violating process without the tunneling suppression factor, it may choose not to. Suppose that cross section at energy $E$ is $\sigma_{EW}(E, {\Delta n}=0)$. Let the total $q_Lq_L$ electroweak cross section at energy $E$ be $\sigma_T (E) = \sigma_{EW}(E,{\Delta n}=0)+\sigma(E,{\Delta n}=\pm 1)$, where we ignore the other (${\Delta n} \ne 0, \pm 1$) contributions for simplification. Let the fraction of the ($B+L$)-violating processes among all electroweak processes be 
\begin{equation}
\label{kappa}
\kappa (E)= \frac{\sigma(E,{\Delta n}=\pm 1)}{\sigma_T(E)}\lesssim \frac{\sigma(E,{\Delta n}=\pm 1)}{\sigma_{EW} (E, {\Delta n}=0)} 
\end{equation}
Now $\kappa=0$ within a band gap (no Bloch wave solution), while it is "tunneling" unsuppressed within a pass band, so we expect $\kappa \lesssim 1$ at energies around and above the sphaleron energy. Averaging over a set of pass bands yields $\kappa <1$, even for energies above $E_{sph}$. To guesstimate the value of $\kappa$,  let us compare the ($B+L$)-violating process (\ref{qqBL-1}) to a $\Delta n=0$ scattering with the same $X$ as in process (\ref{qqBL-1}), i.e., $q_Lq_L \rightarrow qq X$. We see that the extra leptons and quarks ${\bar l}_e  {\bar l}_{\mu}  {\bar l}_{\tau} {\bar q}{\bar q}{\bar q}{\bar q}{\bar q}{\bar q}{\bar q}$ in (\ref{qqBL-1}) have a threshold of about 20 GeV only, so phase space puts little constraint on the process (\ref{qqBL-1}) with energies above $9$ TeV. Also, the ($B+L$)-violating part of the process (\ref{qqBL-1}) is non-perturbative; so, besides the tunneling suppression factor that disappears for $E>E_{sph}$, there are no powers of coupling factors that would have suppressed its rate. So we believe that  $\kappa <1$, but not exponentially small.
It will be very important to find out whether there are other factors that will make $\kappa \ll 1$.

Let us now estimate the ($B+L$)-violating event number at LHC at $E(pp)=14$ TeV. Note that there is a $F_3=1/8$ chance to observe 3 same sign charged leptons instead of one or more neutrinos in such processes.
Let the total inelastic pp cross-section at 14 TeV be $\sigma (pp)\sim 80$ mb (millibarns) (see e.g., Ref\cite{Aad2011}). Taking $f(14 \mbox{TeV}, 9 \mbox{TeV})\sim 10^{-8}$ (which is a crude but conservative estimate), and an integrated luminosity of $L_{pp}=$ 3000 $fb^{-1}$ (inverse femtobarns), we expect the number of ($B+L$)-violating events with 3 same sign charged leptons to be (units in barn and inverse barn) :
\begin{align}
&\mbox{Integrated event number} \nonumber\\
&\sim \sigma(pp)\cdot  f(14, 9) \cdot F_{EW} \cdot \kappa \cdot F_3 \cdot L_{pp} \nonumber\\
&\sim (80 \times 10^{-3}b)(10^{-8})(10^{-2})(10^{-2}/3) (\frac{1}{8})(3000 \times 10^{15}b^{-1}) \nonumber\\
& \sim  10^{4}
\end{align}
where, for lack of better information, we naively take $F_{EW} \sim 10^{-2}$ and the fraction $\kappa$ in Eq.(\ref{kappa}) to be $\kappa \sim 10^{-2}/3$. Crudely, we guestimate the number of such ($B+L$)-violating events to be $10^{4 \pm 2}$ in the LHC 14 TeV run.  Clearly a better understanding of these and other possible factors is important.
There is no question that increasing the pp energy $E(pp)$ by just a few TeV above 14 TeV will increase the event rate by more than an order of magnitude.

Because the quark momenta are at the tail end of the parton distribution functions, we do expect $f(14  \mbox{TeV}, E_c(q_Lq_L))$ to be substantially larger than $f(13  \mbox{TeV}, E_c(q_Lq_L))$, may be by as much as an order of magnitude.  Other factors, such as the poorly understood $F_{EW}$ and $\kappa$,  probably vary much less  as $E(pp)$ increases from 13 TeV to 14 TeV; so they drop out in the ratio. As a result, the ratio of the rates of the ($B+L$)-violating processes may be close to $f(14, 9)/f(13, 9) \gg 1$. Such processes may be detected by a careful comparison of the data from 13 TeV with the data from 14 TeV. 


\subsection{Remarks}

A couple of remarks on the phenomenology are in order :

$\bullet$ As we go to higher energies in a proton-proton collider, larger $\Delta n=-1-K$ processes become possible,
$$ q_L + q_L \rightarrow {\bar l}_e  {\bar l}_{\mu}  {\bar l}_{\tau} {\bar q}{\bar q}{\bar q}{\bar q}{\bar q}{\bar q}{\bar q} +[{\bar l}_e  {\bar l}_{\mu}  {\bar l}_{\tau} {\bar q}{\bar q}{\bar q}{\bar q}{\bar q}{\bar q}{\bar q}{\bar q}{\bar q}]^{K}+ X.$$
For example, at $E(pp)=100$ TeV,  the fraction $f(E(pp), |n| E_{sph})$ for say $|\Delta n| \ge 4$ will have reasonable values for   $q_Lq_L$ scatterings.  In this case, a ($B+L$)-violating process that produces more than a dozen same sign leptons and a dozen of (anti-)b-quarks is quite feasible.


One way to estimate the rate is via a sphaleron with CS number $\frac{n}{2}$.
 There are sphalerons with $N=\frac{n}{2} \ne \frac{1}{2}$, where $E(n/2)_{sph} > nE_{sph}$.
By replacing $U(\varphi)$ (\ref{ours}) by $U(n\varphi)$, the spherically symmetric solution will have topological number $\frac{n}{2}$ while its energy is greater than $nE_{sph}$. Lower energy multi-sphaleron solutions are in general only axially symmetric even without $U(1)_Y$ ($g'=0$) \cite{Kleihaus1994,Kleihaus1994a}. These multi-sphalerons are also repulsive at $m_H=125$ GeV, that is $E(n/2)_{sph} > n E_{sph}$ and turning on $g'$ does not change this feature. There is a direction in the field space where a periodic potential with these $N$-sphalerons separating the vacua exists, so the Bloch wave physics should also apply.
Typically, we expect such a process to become effective when the energy is close to or above that of a multi-sphaleron, $E(n/2)_{sph} \ge n E_{sph}$.
To go from $n=0$ to $n=2$ with some probability, the $q_Lq_L$ energy must be at least 18 TeV.

$\bullet$ If a whole proton can fit inside a sphaleron, then we should consider the total proton-proton energy instead of the energy of its constituents in a ($B+L$)-violating process. However, the typical size of a sphaleron is $m_W^{-1}$ while the size of a proton is $R_P \sim 350 m_W^{-1}$, so sphaleron mediated baryon-number violating processes can take place only for the elementary constituents inside the proton, namely the left-handed quarks as described above. However, this point is not as obvious as it seems.
 
As we have proposed that baryon-lepton number violating tunneling is not suppressed for some specific low energies, and as AKY shows that the size of the to-be-formed sphaleron (i.e., with $0< n \ll 1/2$) at low energies can have a much bigger size, it is interesting to take a closer look on this issue. Following Eq.(\ref{alpha1}), we see that the size of gauge field part of a  to-be-formed sphaleron is given by $\alpha^{-1} \ge m_W^{-1}$, where $\alpha$ is given by Eq.(\ref{eqn:alpha}). Demanding that $\alpha^{-1} \ge R_P$, we find that $n< 10^{-4}$, which means the pp center-of-mass energy is below the lowest band energy, $E < 4$ GeV. So we cannot fit a proton into a to-be-formed sphaleron to generate a ($B+L$)-violating process.


\section{Summary and Remarks}

In this paper, we take advantage of the discrete symmetry of the periodic sphaleron potential to estimate the rate of the ($B+L$)-violating processes in the electroweak theory.
The sphaleron energy $E_{sph}=9.11$ TeV (or 9.0 TeV for phenomenology) measures the height of the potential barrier that separates vacua with different ($B+L$) numbers.  We write down the effective one-dimensional time-independent Schr\"{o}dinger equation for this periodic potential where the coordinate is essentially $\mu$, which is closely related to the CS number $n$ (see Eq.(\ref{nmur})). The choice of $\mu$ yields a constant mass in the kinetic term. It is then straightforward to find the Bloch wave function and the conducting (pass) bands and their widths. 
When the quark-quark energy in a proton-proton collision is around or above $E_{sph}$, such ($B+L$)-violating processes are no longer exponentially suppressed via tunneling. A crude estimate suggests that LHC at 14 TeV has a good chance of detecting them, as some of these events will have 3 same sign leptons plus 3 b-quarks.

Since the quark-quark energies close to but below $E_{sph}$ are at the tail end of the parton distributions for proton-proton energies at 13 TeV, we expect a substantially higher rate of ($B+L$)-violating processes at 14 TeV than at 13 TeV, so a comparison of the events at these two energies will be very interesting. If observed, they can provide a probe into other physics as well. For example, $E_{sph}$ will be shifted if there are 2 Higgs doublets, or via quantum corrections or modified Higgs potential, topics not discussed in this paper. 

Bloch wave function is a special (particularly elegant) example of resonant tunneling: that is, tunneling can be very efficient when the resonance condition is satisfied. Resonant tunneling can happen only when there are 3 or more minima. This is very well understood in quantum mechanics, but much less so in quantum field theory. There is a very puzzling property in Helium-3 superfluidity, which has quite a number of distinct phases. If one starts in the A phase (while the actual ground state is in the B-phase) and tunes the temperature, pressure and/or magnetic field slightly, it will go to the B phase via a first order phase transition. As the field theory of He-3 superfluid is well studied, theoretical calculation shows that such a phase transition will take a minimum $10^{20,000}$ years to happen \cite{Bailin1980}. Yet, in the laboratory, it happens readily. Now, being a superfluid, there are no impurities inside the sample; furthermore, the A$\rightarrow$B phase transition must happen in the bulk of the sample away from the container wall, as wall effect stabilizes the He-3 to stay in the A phase. So there is nothing to seed nucleation bubble. Now, this puzzle may be explained by resonant tunneling \cite{Tye2011}.  Here, the ($B+L$)-violating process may be another case where we can directly check the resonant tunneling phenomenon in field theory in the laboratory.
 
 Our analysis, in particular the evaluation of the mass $m$, is based on two different approaches in the static approximation, namely that of the Manton and the AKY approaches. These two methods take different approximations to the actual problem. The differences in the band structures give us a rough idea of the uncertainties involved. Clearly a better evaluation is desirable. This probably requires a time-dependent treatment of the sphaleron potential and mass. 

Note that we have assumed the simplest Higgs potential in Eq.(\ref{lagran}), with a quadratic plus a quartic term in $\Phi$. Since only the Higgs  vacuum expectation value and the Higgs mass have been measured so far, the Higgs potential may take other forms or there may be more than one Higgs doublet. This will change the sphaleron energy $E_{sph}$ (as well as the sphaleron potential shape) which in turn would change the threshold energy where the ($B+L$)-violating processes becomes unsuppressed. So the observation of such events at LHC will provide valuable information about the Higgs potential itself as well as the Higgs self-couplings. 
 
A better understanding of the ($B+L$)-violating processes at the colliders will surely improve our understanding of the ($B+L$)-violating processes during the electroweak phase transition in the early universe. We can also assess the possibility whether such processes will enhance the anti-proton to proton ratio in our universe. 

A typical energy released in a nuclear reaction is of order MeV, while the energies released in the ($B+L$)-violating processes are of order of GeV (via the annihilation of baryons with the anti-baryons produced), roughly a factor of thousand bigger. Once we understand the physics better, we may find a way to take advantage of this release of energy.

\section*{Acknowledgments}
We thank Che-Ting Chan, John Ellis, Jan Hajer, Ho Tat Lam, Ying-Ying Li, Tao Liu, Kirill Prokofiev,  John F.H. Shiu, Sichun Sun, Dick Talman and Yanjun Tu for discussions. The work is supported by grants HKUST4/CRF/13G and the GRF 16305414 issued by the Research Grants Council (RGC) of Hong Kong. 
 
\appendix

\section{The function $f(r)$ and $h(r)$}
We present in Fig. 9 the numerical results of the profile-function $f(r)$ and $h(r)$ of the gauge field and Higgs field respectively as defined in Eq.(\ref{ours}).

\begin{figure}
 \begin{center}
  \includegraphics[scale=.5]{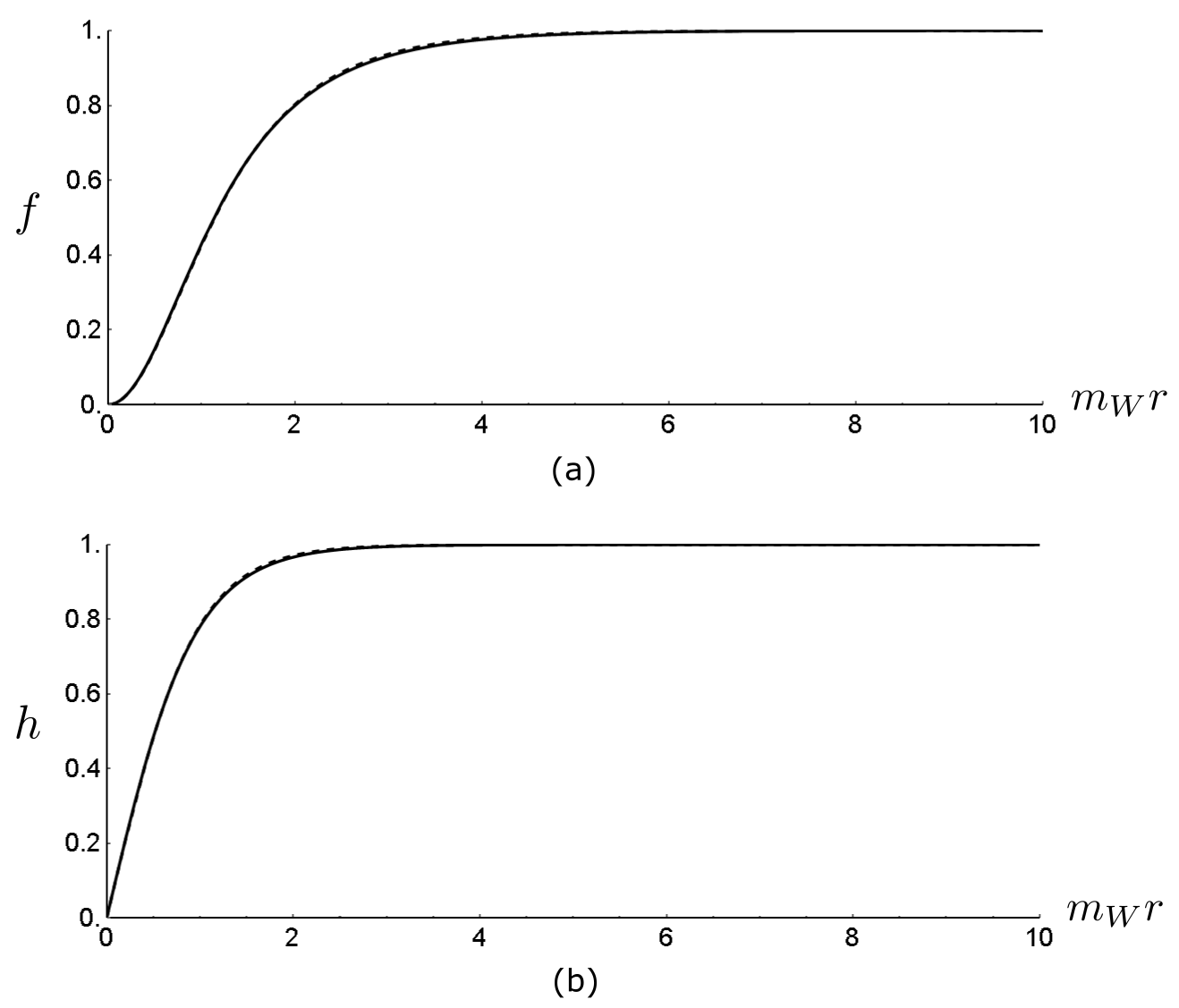}
  \caption{(a) Numerical result (solid line) of $f(r)$ and the approximate function (dashed line) $f(r) \approx 1-\mbox{sech}(1.154 m_W r)$. (b) Numerical result (solid line) of $h(r)$ and the approximate function (dashed line) $h(r) \approx \tanh (1.056 m_W r)$. The numerical forms and the approximate functions almost overlap.} 
 \end{center}\label{fhr9}
\end{figure}

Following the asymptotic expansion of $f(r)$ and $h(r)$ at the origin, one can approximate them by 
\begin{equation}
f(r) \approx 1-\mbox{sech}(a m_W r), \quad h(r) \approx \tanh (b m_W r)
\end{equation}
and then determine $a=1.154$ and $b=1.056$ by minimizing the energy $V_M(\mu)$ (\ref{Vmu}) at $\mu=\pi/2$.
See Fig. 9 
for comparison.

\FloatBarrier
\bibliographystyle{utphys}
\bibliography{References}

\end{document}